\documentclass{JHEP3}
\usepackage{amsmath,amsfonts,bbm,euscript, amssymb}
\usepackage{epsfig,cite}
\usepackage{graphicx}
\usepackage{psfrag}
\newcommand{\be}{\begin{equation}}
\newcommand{\ee}{\end{equation}}
\newcommand\beq{\begin{eqnarray}}
\newcommand\eeq{\end{eqnarray}}

\newcommand{\GeV}{{\rm ~GeV }}
\newcommand{\TeV}{{\rm ~TeV }}

\newcommand{\mvh}{m_{vh }}

\usepackage{axodraw}

\title{Phenomenology of Hidden Valleys at Hadron Colliders}
\author{}
\author{Tao Han$^1$, Zongguo Si$^2$, Kathryn M. Zurek$^1$, Matthew J. Strassler$^3$\\
$^1$Department of Physics, University of Wisconsin, Madison, WI 53706, USA \\
$^2$Department of Physics, Shandong University, Jinan, Shandong 250100, P.R. China \\
$^3$Department of Physics, Rutgers University, Piscataway, NJ 08854}
\preprint{MADPH-07-1502, SDU-HEP-071201, RUNHETC-2007-31}
\abstract{We study the phenomenology of, and search techniques for, a class of  ``Hidden Valleys." These models are characterized by low mass (well below a TeV) bound states resulting from a confining gauge interaction 
in a hidden sector; the states include a spin-one resonance that can decay to lepton pairs. Assuming that the hidden sector communicates to the Standard Model (SM) 
through TeV suppressed operators, taking into account the constraint from the $Z$ pole physics at LEP,  
searches at Tevatron may be difficult in the particular class of Hidden Valleys we consider, so that we concentrate on the searches at the LHC.
Hidden Valley events are characterized by high multiplicities of jets and leptons in the final state.  
Depending on the scale  of confinement in the hidden sector, 
the events are typically more spherical, with lower thrust and higher incidences of 
isolated leptons, than those from the SM background processes.  Most notably, high cluster invariant mass
and very narrow, low mass resonances in lepton pairs are the key observables to identify the signal.  We use these characteristics to develop a set of cuts to separate the Hidden Valley from SM, and show that with these cuts LHC has a significant reach in the parameter space. Our strategies are quite general and should apply well beyond the particular class of models studied here.}

\begin{document}
 
\section{Introduction}

In the anticipation of major discovery at the Large Hadron Collider (LHC) for physics beyond 
the standard model (SM)
at the TeV scale, it is prudent to keep in mind that the high energy frontier may open up
the possibility for new physics with non-conventional forms. Many theoretical
models include a ``hidden sector" that couples to the SM weakly.
The hidden sector may communicate to the standard model through a heavy mediator (typically with a mass scale
 of TeV or higher), charged under both the SM and hidden sector. TeV scale particle physics model building is replete with examples of such hidden sectors.  These include gravity mediated SUSY breaking \cite{Chamseddine:1982jx,Barbieri:1982eh,Hall:1983iz}, gauge mediated SUSY breaking \cite{Dine:1993yw,Dine:1994vc,Dine:1995ag}, Twin Higgs models \cite{Chacko:2005pe}, milli-charged hidden sector dark matter \cite{Feldman:2007wj,Cheung:2007ut} or unparticle physics \cite{Georgi:2007ek,Georgi:2007si}.  String models also generate hidden sector matter \cite{Cvetic:2002qa,ArkaniHamed:2005yv}. 

We consider in this paper a class of hidden sector models which are characterized by a new confining gauge interaction in the hidden sector.   The dynamics in the hidden sector is set by the confinement scale, which introduces a mass gap into the theory.   The addition of new quarks into the hidden sector will give rise to a 
variety of hidden bound states (as in QCD), with presumably low mass much below 1 TeV.
This class of models, introduced in \cite{Strassler:2006im}, in an example of a ``confining Hidden Valley.''  The term ``Hidden Valley'' refers to the presence of the low mass ``valley'' ($v$-)states ($v$-quarks and  $v$-hadrons in the present case), which can only be observed by passing through or over a barrier separating the hidden sector from the standard model sector.  
The novel phenomenology of the Hidden Valley arises because of the low mass of the $v$-particles and the nontrivial dynamics in the hidden sector. 
The  masses  of the $v$-hadrons are determined by either the $v$-quark mass or the 
confinement scale, whichever is larger. The lower bound on the $v$-hadron's mass 
is derived by requiring that it decay before BBN; the precise constraint thus depends on the dimension of the operator mediating communication between the hidden sector and the standard model, but it is typically sub-GeV.   When the mass of the bound state is below 10 MeV, on the other hand, the dominant constraints may be derived from astrophysics and cosmology.

%
 
Similar to the mediation of SUSY breaking, the effects of a Hidden Valley on our SM observable sector are 
crucially determined by the nature of the mediators \cite{Strassler:2006im,Strassler:2006ri,Strassler:2006qa}.  
Given the presumably large gap between the mass of $v$-sector particles and the heavy mediators, 
there is effectively a barrier (like a mountain pass) between the SM and Hidden Valley which must be surmounted in order to produce the valley particles via the SM processes. 
When the mediator is integrated out, an effective  operator is generically of the form
\be
\label{highdim}
g_v g_{SM}^{} \frac{{\cal O}_v{\cal O}_{\rm SM}}{M^k} ,
\ee
where $g_v$ and  $g_{SM}^{}$ are the mediator couplings to the two sectors, 
$M$ the mass of the mediator 
and the power of the suppression, $k$, depends on the mediator and the nature of the hidden sector.  Examples of possible mediators are $Z'$'s, Higgs sectors, messenger fermions 
which connect the gluons from the hidden and standard model sectors, or gravitons from 
Randall-Sundrum or large extra dimensions. 
On the other hand, if the mediator scale $M$ is accessible by the experiments, one could thus expect
enhanced effects. 



Once the $v$-sector quarks are produced through the mediator, they confine into $v$-hadrons.  If the $v$-hadrons are much lighter than the mediator, many of them are produced in the $v$-sector hadronization.
Provided the mass gap is not too low and the hidden sector resonances not too light, these $v$-hadrons decay 
back through the mediator to combinations of leptons and quarks which, when summed together, are standard model neutral since the $v$-hadron itself carries no standard model charges. Thus $v$-hadrons, including the ones that will be the focus of our study, may decay to neutral pairs of leptons, heavy quarks, or light quark jets.  In particular, the decay of spin-one resonances tends to be democratic among standard model fermions. 

To be concrete and to gain a basic understanding of the features and search strategies at a collider, we choose a particular model for the hidden sector and mediator.  We take a $Z'$ mediator and a hidden sector with one light quark, which has narrow spin-zero and spin-one resonances and no absolutely stable $v$-hadrons.  Another example of such a model has been studied elsewhere \cite{unpubMrennaSkandsMJS} though no signal-to-background studies have been performed.  Many of the features we describe here will be transferrable to a broader class of models with a confining gauge group in the hidden sector.  We choose to focus on the region of parameter space in this model where the $v$-hadrons decay promptly back to SM particles, which was shown in \cite{Strassler:2006ri} to occur for
\be
\mvh \gtrsim 30\ \mbox{ GeV}.
\label{masses}
\ee
Lighter mass $v$-hadrons will either decay with a displaced vertex in the detector or will result in missing energy, leading to qualitatively different signals which are not the focus of this study. 
The general features we will study for the prompt decays are independent, in a qualitative way, of the mass of the $v$-hadron and the specific nature of the mediators. 
We will demonstrate that these new confinement dynamics may lead to quite distinctive features
that are characteristically different from the SM expectations, and demonstrate the ways in which the hidden sector strong dynamics may be separated from SM QCD.

It should be kept in mind that Hidden Valley phenomenology can be diverse far beyond the simple setting laid out above. It 
depends on the size of the mass gap, the nature of the mediator, and the matter content of the hidden sector, all of which enter the effective operator, Eq.~(\ref{highdim}).  For example, in the limit that the mass gap is taken to zero and the quarks are massless, we recover a model of scale invariance, similar to Unparticles 
 \cite{Georgi:2007ek,Georgi:2007si}.
On the other hand, if all the quarks are much heavier than the confinement scale,  $v$-quarkonium states can play an important role in the physics \cite{Strassler:2006im}, and if some of the heavy $v$-quarks also carry standard model charges, ``quirk'' phenomenology results \cite{Kang:2008ea,Burdman:2008ek}.  
In some models, certain of the bound states may be stable leading to large missing energy signals.  Some of the states may be quasi long-lived, decaying a macroscopic distance away from the interaction region giving rise to the displaced vertex discussed in \cite{Strassler:2006im,Strassler:2006ri}. 
With a new Higgs mediator that mixes with the SM Higgs boson, Hidden Valleys may give rise to novel search techniques for the Higgs through highly unusual decay patterns, some of which may be implemented at the Tevatron and LHC. Clearly there is a rich and broad phenomenology which remains to be explored.

The rest of paper is organized as follows. In Sec.~2, we outline a general scenario of a 
Hidden Valley model, reiterate the current bounds, 
and calculate the typical production rates for the signal at the LHC. 
Our treatment for the $v$-quark hadronization is also described. In Sec.~3, we systematically explore a method for separating Hidden Valley signals from the QCD background which relies on distinguishing the unique Hidden Valley event shapes; we discuss the predominant backgrounds to our signal.  We conclude in Sec.~4.

\section{The Model And The Hidden Valley Particle  Production}

\subsection{Model description}

We utilize the model of \cite{Strassler:2006im} in computing production and decay.  In this model, SM
fermions such as $q$ and $\bar{q}$ annihilate through the $Z'$ mediator to hidden sector quarks $v$
and $\bar{v}$.  The charges are repeated for convenience in Table~\ref{charge}.  We choose the simplest hidden sector content possible, a single light $v$-quark, which we denote $U$ in the table,  though an additional heavy hidden sector quark (not relevant for the phenomenology discussed here) must be added to make the theory anomaly free.  
  
\TABLE[t]{\label{charge}
\begin{tabular}{|c||c|c|c|c|c|c||c|c||c|c|| }
\hline
& $q_i$ & $\bar{u}_i$ & $\bar{d}_i$ & $l_i$ & $e_i^+$ & $N_i$ & $U$ & $\bar{U}$ & $H$ & $\phi$ \\ \hline
$U(1)_\chi$ & $\frac{-1}{5}$ & $\frac{-1}{5}$ & $\frac{3}{5}$ & $\frac{3}{5}$ & $\frac{-1}{5}$ & $-1$ & $q_+$ & $q_-$ & $ \frac{2}{5}$ & 2 \\ \hline
$SU(\hat{N}_c)$ & 1 & 1 &  1 &  1 &  1 &  1 & $\check{N}$ & $\bar{\check{N}}$ & 1 & 1 \\ \hline
\end{tabular}
\caption{Charges of the SM and hidden sector quarks under the mediator gauge group, $U(1)_\chi$, and the hidden sector confining group, $SU(\hat{N}_c)$.  The hidden sector quarks are uncharged under $SU(3)\times SU(2) \times U(1)$ and $q_+ + q_- = -2$. }
}

\FIGURE[t]{\label{fig:feyndia}
\epsfxsize=2.9in \epsfbox{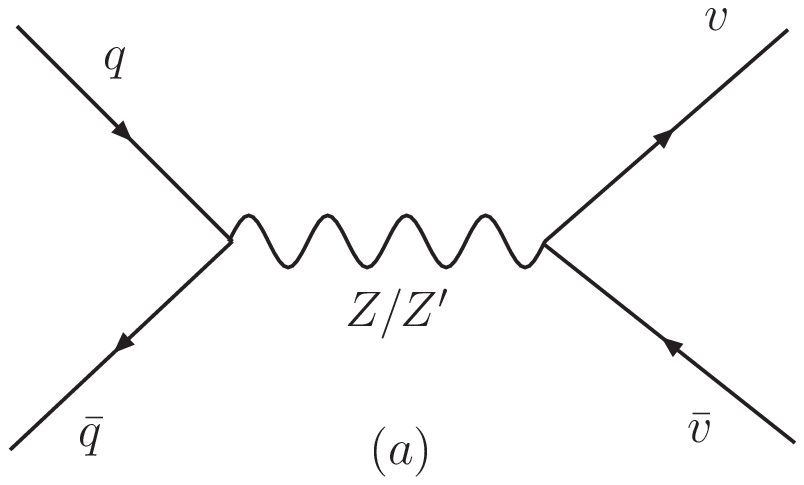}
\epsfxsize=2.9in \epsfbox{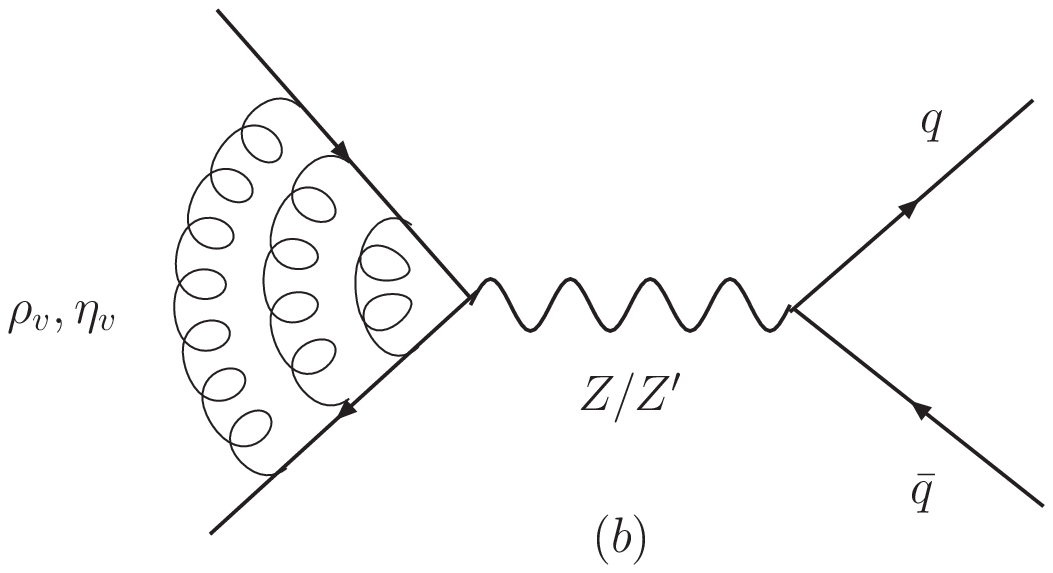}
\caption{(a) Production of  $v$-quarks  and (b) decay  of 
$v$-hadrons,  $\rho_v$ (the vector bound state of $v$-quarks), $\eta_v$ (the pseduoscalar bound state).}
} 

Since the $Z$ and $Z'$ are charged under both the hidden and visible Higgs sectors, the $Z$ and $Z'$ mix through mass terms.  As a result, $v$-quarks may be produced through either a $Z$ or $Z'$ mediator, as indicated in Fig.~\ref{fig:feyndia}a.  The most stringent constraints on models of this type arise from the LEP measurements on the $Z$ pole.  We assume that no more than a few events of $v$-quark production will be consistent with the constraints from LEP, so that we require a branching fraction less than $5 \times 10^{-7}$ from decays of $Z$ to $v$-quarks at LEP through $Z-Z'$ mixing.  In particular, the production cross-section is 
\beq
\label{totalcross-section}
\sigma_v  & = &  \frac{\hat{N}_c g'^4}{48 \pi}  \left(\left|R_e^L\right|^2+\left|R_e^R\right|^2\right)\left(Q_v^2+Q_{\bar{v}}^2\right) \nonumber \\
&& \frac{s}{(s - m_{Z'}^2)^2+\Gamma_{Z'}^2 m_{Z'}^2} 
\left(1-4\frac{m_{v}^2}{s} \right)^{1/2}\left(1+2\frac{m_{v}^2}{s}\right),
\eeq 
where $g'$ is the gauge coupling of the $Z'$, $Q_v$ is the charge of the $v$-quark under $U(1)_\chi$, $m_v$ is the mass of the $v$-quark, and $R_f$ includes the effects of $Z-Z'$ mass mixing:
\be
R_f^{L,R} = Q_f^{L,R \mbox{ } Z'} - 2 Q_f^{L,R \mbox{ } Z} Q_H \frac{ m_Z^2}{s-m_Z^2 + i \Gamma_Z m_Z}.
\ee 
This is to be compared against the standard model total cross-section: $e^+ e^- \rightarrow Z \rightarrow anything$, $\sigma_{SM} = g_Z^2 m_Z/ 4 \Gamma_Z $. Requiring $\sigma_v / \sigma_{SM} < 5 \times 10^{-7}$ imposes a bound $m_{Z'}/g' \gtrsim 7 \TeV$.  In the remainder of this paper, we take 
\be
\label{para}
m_{Z'} = 1\  {\TeV},\quad m_{Z'}/g' = 7\  {\TeV},\quad \hat{N}_c = 3. 
\ee
Later we generalize the parameters to determine the LHC reach.

\subsection{On-threshold production: The $v$-onia production}

\FIGURE[t]{
\epsfxsize=3.5in \epsfbox{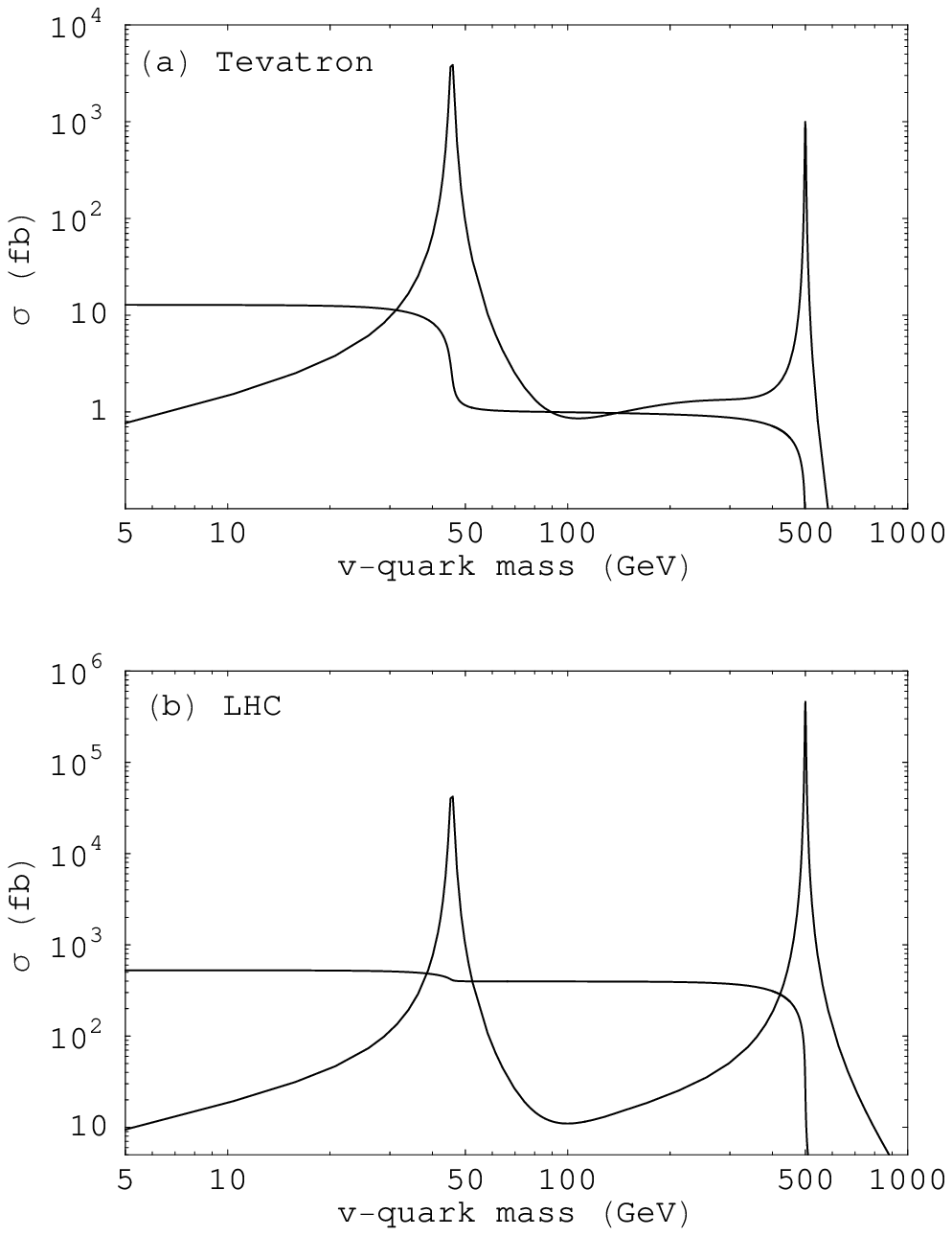}
\caption{Total cross section of  $v$-hadron production versus the $v$-quark mass $m_v$ at (a) the Tevatron
and (b) the LHC. The peaked curves are on-threshold $v\bar v$ bound-state production, 
the flatter curves open-flavor $v\bar v$ pair production.  
Here and throughout this paper the parameters $m_{Z'} = 1 \mbox{ TeV}$, $m_{Z'}/g' = 7 \mbox{ TeV}$ and $\hat{N}_c = 3$ are chosen.}
\label{fig:tot}
}  

Historically, new bound-states have been discovered on threshold, where $\sqrt{s} = 2 m_q$.  The existence of $c$ and $b$ quarks was gleaned through the observation of the $J/\psi$  and $\Upsilon$  bound-states. 
It is thus natural to consider the $v\bar v$ bound state formation as the signal production mechanism.

For the case of a heavy quark, the calculation of on-threshold bound-state production follows that of quarkonium (see \cite{Barger:1987nn} for details of quarkonium production):
\begin{eqnarray}
\sigma_{thresh} = \int_{4 m_v^2/s}^1 dx_1 \int_{4 m_v^2/s x_1}^1 dx_2 &  \frac{\hat{N}_c}{N_c} \frac{g'^4 \pi \left(\left|R_q^L\right|^2+\left|R_q^R\right|^2\right) \left(Q_v^2+Q_{\bar{v}}^2\right) \left| \phi_{vh} \right|^2 m_{vh}}{(S x_1 x_2 - m_{vh}^2)^2+\Gamma_{Z'}^2 m_{Z'}^2} \delta(S x_1 x_2 - m_{vh}^2) \nonumber\\
 & \times \sum_q 2 f_q(x_1) f_{\bar{q}}(x_2),
\end{eqnarray}
where $S$ is the hadronic center of mass energy ($S x_1 x_2 = s$).  For perturbative quarkonium, the wavefunction $\phi_{vh}$ can be computed analytically in a manner analogous to the hydrogen wavefunction: $\left|\phi_{vh}\right|^2 = \left[\frac{2}{3}\hat{N}_c^{-1} m_v \hat{\alpha}_s(m_v^2)\right]^3/\pi$, where $m_v$ is the $v$-quark mass.  For non-perturbative bound states of $v \bar{v}$, the wave function cannot be computed analytically, but can only be approximated through the relation $\left| F_{V,A} \right|^2 = 8 \hat{N}_c m_{vh} \left| \phi_{vh} \right|^2$, with $F_{V,A}$ the vector and pseudoscalar decay constants. 
The decay constants must be determined either experimentally or by non-perturbative lattice calculations.   

The production cross-section is shown  in Fig.~\ref{fig:tot} as a function of the $v$-quark mass (a) at the Tevatron
and (b) at the LHC energies for a mediator mass $m_{Z'}=1$ TeV, 
assuming $F_{V,A} = m_{vh}^2$ (the result can be easily rescaled for different decay constants).  
We see the large resonant enhancement  at $2m_v\approx m_Z$ or $m_{Z'}$
due to the mediators between the SM fields and the $v$-sector. Off the $Z$ or $Z'$ resonances,
the production cross-sections are typically of the size a few fb at the Tevatron and 10's of fb or smaller
at the LHC. The smallness of the cross section is largely due to the suppression of the $Z'$ propagator.
Given the several orders of magnitude of larger rates for the SM DY production as backgrounds, 
the $v$-onia signal may not be readily observable. We will not pursue this channel further.

\subsection{Open-flavor production and hadronization}

\FIGURE[t]{
\epsfxsize=4in \epsfbox{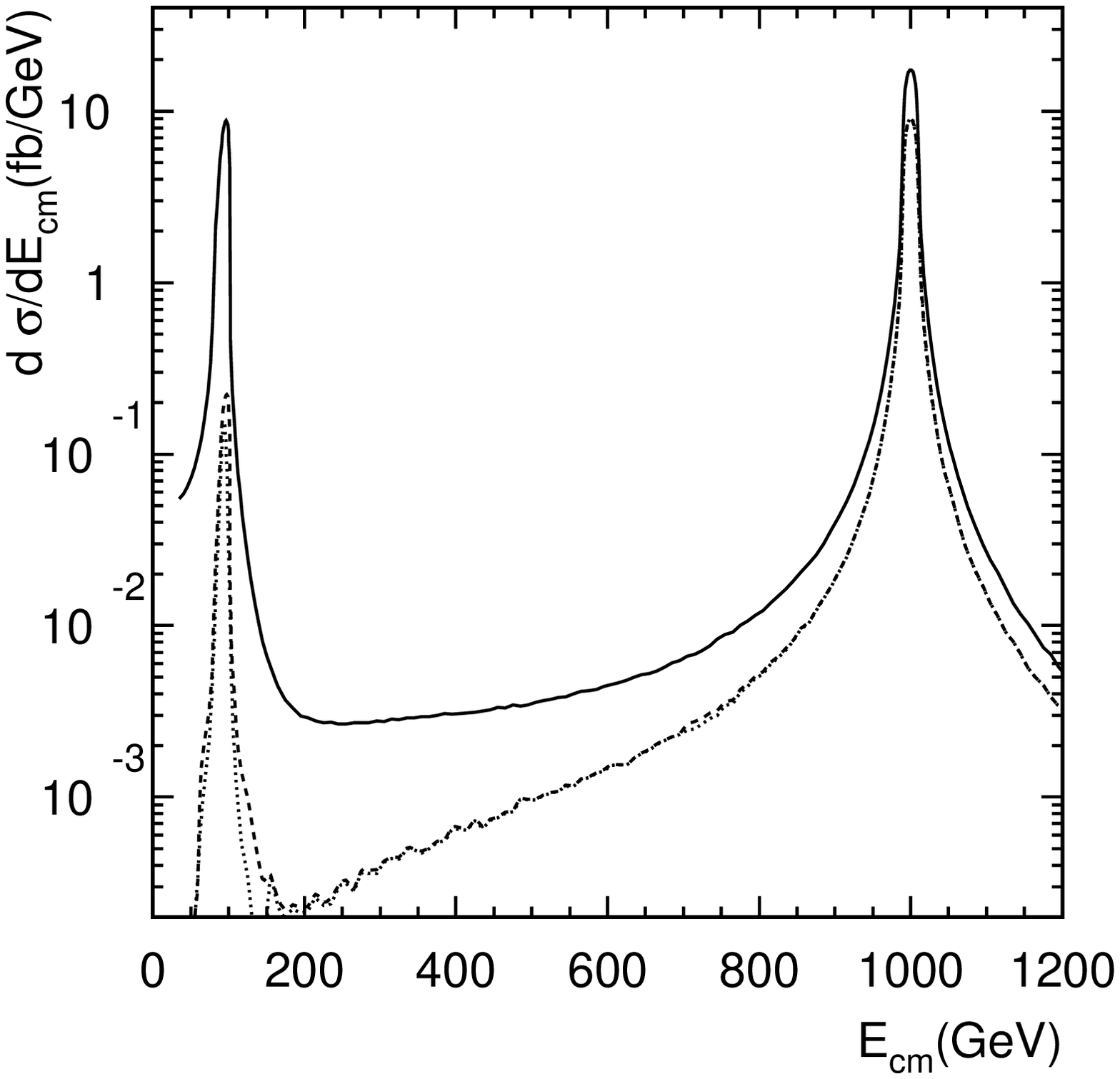}
\caption{Differential cross-sections as a function of partonic c.m.~energy $\sqrt{{s}}$.
The solid curve represents the inclusive cross-section without acceptance cuts, the dashed is
with the trigger requirement in Eq.~(\ref{eq:mu}), and the dotted includes further cuts as  
given in Eqs.~(\ref{eq:tcut}),~(\ref{eq:scut}), and (\ref{eq:mclustercut}).}
\label{csdis}
}
Once produced, the $v$-quarks will hadronize into a ``$v$-color" singlet state, as in QCD. 
If they are produced near the threshold $2 m_v$, it is likely that the $v\bar v$ will form a 
single bound state, $\rho_v$ or $\eta_v$ (the vector and pseduoscalar bound states of $v$-quarks, respectively),  as we discussed in the previous section with production rates on-threshold shown in Fig.~\ref{fig:tot}.
If produced  well above threshold, they form multiple $v$-color singlet bound-states, leading to
possible $v$-jets.  We make use of the production cross-section Eq.~(\ref{totalcross-section}), 
appropriately modified with parton distribution functions 
at a hadron collider.  The total open-flavor cross-section is shown in Fig.~\ref{fig:tot} 
as a function of $m_{v}$ (the flatter curves). As expected, the production rate is much larger than the
bound state production when $2m_v \ne m_Z$ or $m_{Z'}$.
The $v$-quarks are now being produced through a resonant decay of the mediator 
that couples the $q\bar q$ in the initial state to $v\bar v$ in the final state. 
The cross section is nearly constant up to the kinematical limit $m_v \approx m_Z/2,\  m_{Z'}/2$,
and the step-function nature reflects the fact crossing over the mediator thresholds.
Once again, we see that the contribution from a 1 TeV $Z'$ is significantly larger than that
from the $Z$ at the LHC, while it is opposite at the Tevatron due to the limited parton c.m.~energy.
The cross section scales with the model parameters as
\begin{eqnarray}
 \sigma \propto \left\{
 \begin{array}{cc}
 g'^2/m_{Z'} \     &  {\rm  at\  the}\  Z'\  {\rm resonance,}\\
 (g'/m_{Z'})^4 \  &  {\rm  off\ the}\ Z' {\rm resonance,\  even\ at\  the}\  Z\  {\rm resonance}.
\end{array}
\right.
\label{behave}
\end{eqnarray}
Given the expected integrated luminosity of a few fb$^{-1}$ at the Tevatron, the signal rate would
be rather low, and it will be thus a challenge to observe the Hidden Valley effects. 
We will henceforth focus on the search at the LHC.

At the LHC, the majority of the contribution comes from $Z'$, three times as much as that from $Z$.
The characteristic kinematics to look for the $v\bar v$ is at the invariant mass peak of the
sub-process, as shown in Fig.~\ref{csdis}, where a 1 TeV $Z'$ and mixing with the SM $Z$ are evident. 
However, whatever the mediator turns out to be, it is a character with known properties. We thus
will make use of the knowledge about it such as the mass, width and typical couplings to the
SM particle etc. to deduce the existence of the Hidden Valley signal. To simplify the analyses,
we will devise an initial cut 
to center on the new mediator mass, as most of the rate is produced there.
For definiteness  and convenience for the presentation, we choose
\beq
\label{MZZcut}
&& m_{Z} - 50\ {\rm GeV} < \sqrt{s} < m_{Z} + 50\ {\rm GeV}\quad  {\rm for}\  Z\ {\rm mediator}\\
&& m_{Z'} - 100\ {\rm GeV} < \sqrt{s} < m_{Z'} + 100\ {\rm GeV}\ {\rm for}\  Z'\ {\rm mediator}.
\label{MZZpcut}
\eeq
This pre-selection condition will effectively separate the background considerations
as well.  For the $Z'$, this requirement is essentially a lower cut on the energetics, given the steeply falling parton luminosities.

To incorporate the hadronization process for the pair-produced $v$ quarks, we write 
\begin{equation}
\sigma_{vh} = \int d x_1\, d x_2 \, d\hat{\sigma}(s)_{q\bar{q}\rightarrow v\bar{v}} \,
P_{v\bar{v}\rightarrow h's},
\end{equation}
where $ d\hat{\sigma}(s)_{q\bar{q}\rightarrow v\bar{v}}$ is differential cross section for the partonic process
$q\bar{q}\rightarrow v\bar{v}$, and $P_{v\bar{v}\rightarrow h's}$ is a production probability 
for the initial $v\bar{v}$-quark pair to transform into $v$-hadrons,
which is described by hadronization models, such as
Lund Fragmentation Model \cite{Sjostrand:2006za}, 
Webber Cluster Model \cite{Corcella:2000bw}, and Quark 
Combination Model (QCM)\cite{Xie:1988wi, Wang:1996jy, Si:1997zs, Si:1997rp, Si:1997ux} etc. 
Here we adopt a modified QCM where the baryon production is neglected. 
Within this model, $N$-pairs of new $v\bar{v}$ can be produced 
from the vacuum just as standard model light quark pairs.
$N$ is assumed to satisfy a Poisson Distribution:
\begin{equation}
\label{fir}
P(\langle N \rangle, N-1) = {\langle N \rangle^{N-1} \over (N-1)!}\  e^{-\langle N\rangle}
\end{equation}
where $\langle N \rangle$ is the average number of quark pairs.
According to Ref.~\cite{Xie:1988wi, Si:1997zs}, $\langle N\rangle$ is parameterized by
\begin{equation}
\langle N \rangle=
\sqrt{\alpha^{2}+\beta \sqrt{s}} -\alpha-1, ~~~~~\\
\alpha=\beta m_v -{{1} \over {4}},
\end{equation}
where $\beta$ is a free parameter\footnote{In this paper, all of the parameters used 
in the hadronization model are fixed by data from $e^+e^-$ between 
$\sqrt s=$10 and 91 GeV \cite{Si:1997zs}. This has been widely used in the literature.},
and $m_v=m_{vh}/2$ is the constituent $v$-quark mass.
Neglecting the $v$-baryon production, these $N$ $v$-quark pairs are assumed to form $N$ $v$-mesons. 
Typical probability distributions of
 multiplicities are shown in Figs.~\ref{multiplicity}a and \ref{multiplicity}b 
 with a $Z$ and $Z'$ mediator respectively, for a various values of  $m_{vh}$.
 We see the qualitative difference between these two regions.
 For a large values of $\mvh$,  the value of $N\times \mvh$ is roughly mass of the mediator,
 while for a small values of $\mvh$, $N$ increases approximately logarithmically and the value
 of $N\times \mvh$ is less than the mass of the mediator,
 with the rest of the energy from the mediator decay being carried away as kinetic energy
 of the light $v$-mesons.

\FIGURE[t]{
\epsfxsize=2.9in \epsfbox{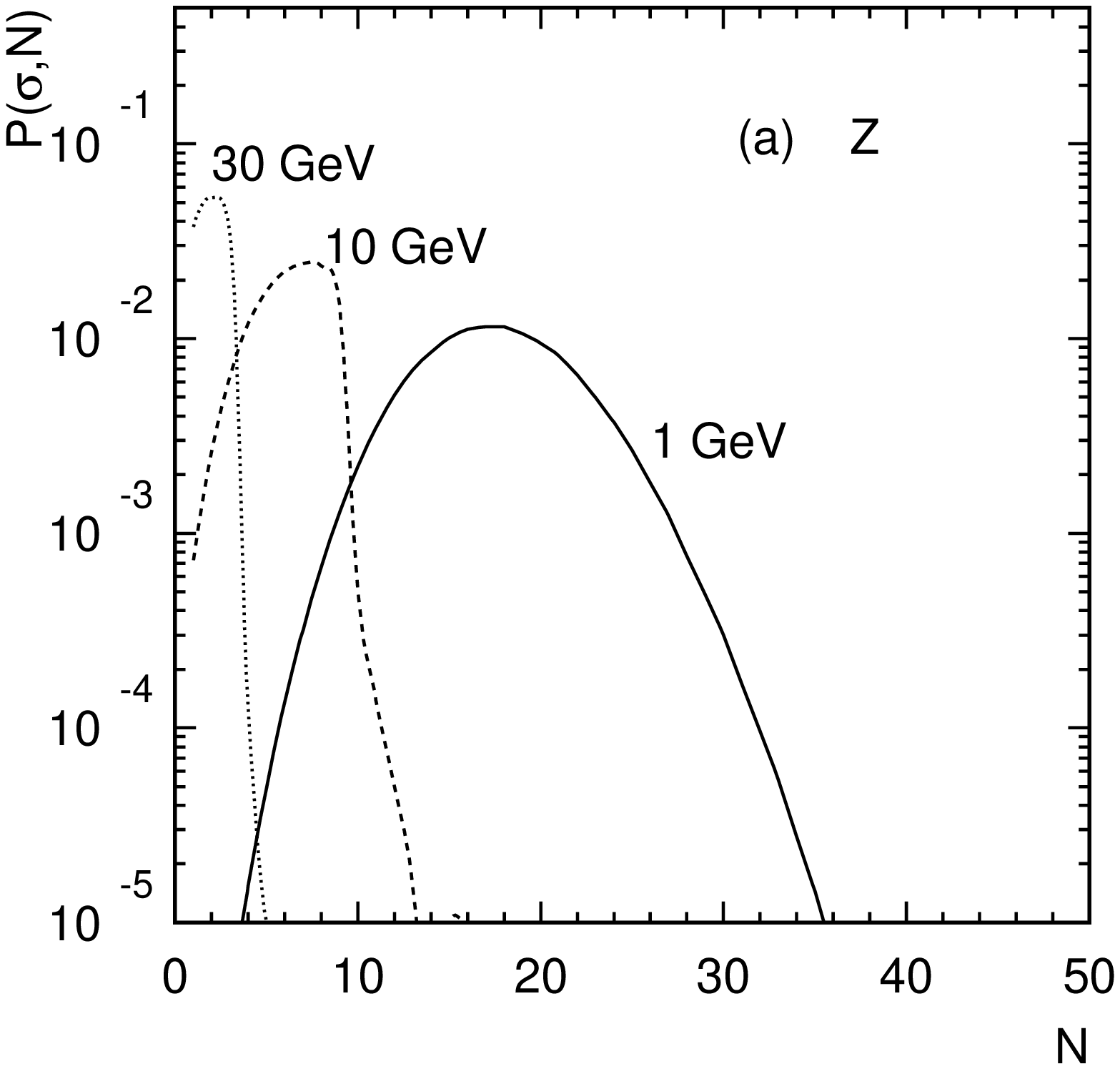}\epsfxsize=2.9in \epsfbox{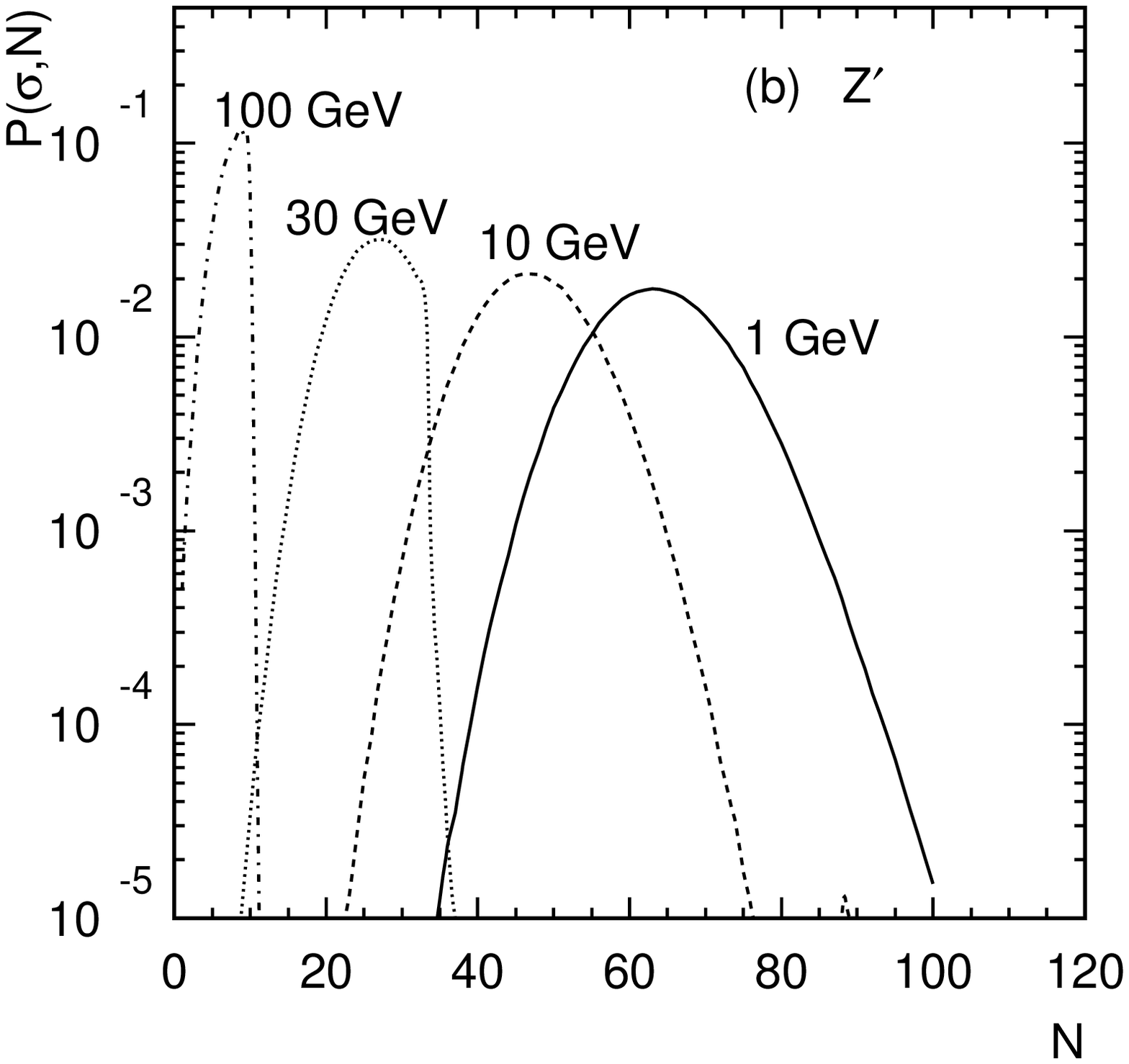}
\caption{Probability distribution of an event containing $N$ $v$-hadrons (a) via a $Z$ mediator and (b) via a $Z'$ mediator.}
\label{multiplicity}
}  

In order to determine the momentum of these $N$ $v$-mesons,
we  simply adopt the widely used  Longitudinal Phase Space
Approximation~(LPSA) which is equivalent to a
constant distribution of rapidity. The hypothesis of the longitudinal phase space
approximation leads to a probability distribution
which is identical to that of a grand canonical ensemble
of non-interacting particles. This approximation is used
to describe successfully the experimental data in
$e^+e^-$ annihilation, e.g. \cite{Kurihara:1988gw}, and is also discussed in many other places, for example
 \cite{TaiTsun:1970nk,Ochs:1981fy,Ochs:1983yd,Richey:1976rf}.  Since a $v$-meson $i$ is
uniformly distributed in rapidity, its rapidity $Y_i$ can be
written as
\begin{equation}
Y_{i}=Z+\xi_{i} Y,
\end{equation}
where $\xi_i$ is a random number between $0$ and $1$;
$Z$ and $Y$ are two arguments which can be determined by
energy-momentum conservation in the initial $v\bar{v}$ system:
\begin{equation}
{\sum \limits_{i=1}^{N}} E_{i} = \sqrt{s},\qquad 
{\sum \limits_{i=1}^{N}} P_{Li} = 0,
\end{equation}
where $E_{i}$ and $P_{Li}$ denote the energy and the longitudinal
momentum of the $i$th $v$-meson. They are
obtained by
\begin {eqnarray}
E_{i} = m _{Ti} {{\exp(Y_{i}) + \exp(-Y_{i})} \over {2}},\ \quad
P_{Li}= m _{Ti} {{\exp(Y_{i}) - \exp(-Y_{i})} \over {2}},
\end{eqnarray}
where the transverse energy 
$m _{Ti} = \sqrt{m_i^2 + {\vec P_{Ti}}^{2}}$,
with $m_{i}$ the mass of the $i$th $v$-meson,
and its transverse momentum $\vec P_{Ti}$ obeying the distribution
\begin{equation}
f({\vec P_{T1}},\ldots,
{\vec P_{TN}}) \propto {\prod \limits_{i=1}^{N}}
\exp(-{{{\vec P_{Ti}}^{2}} \over {\bar{\sigma}^{2}}})\ 
\delta ({\sum \limits_{i=1}^{N}} {\vec P_{Ti}}).
\end{equation}
$\bar{\sigma}$ is a free parameter which we again fix with $e^+e^-$ annihilation data \cite{Si:1997zs}, where $\bar{\sigma}$ is rescaled from the SM value by the ratio $\Lambda_{QCD}/\Lambda_v$, 
where $\Lambda_v$ is the confinement scale in the valley, assumed to be the $v$-hadron mass
 $m_{vh}$ in our treatment.
Using the simplified model described above, we obtain the 4-momenta of the $v$-mesons.

\subsection{$v$-hadron decay}
Each $v$-hadron produced will subsequently decay to SM leptons or hadrons through the diagram shown in Fig.~\ref{fig:feyndia}b.  The branching fractions can be computed from the charges in Table~\ref{charge}, with the assumption that 3/4 of the $v$-hadrons are vectors (which decay democratically to heavy and light flavor alike) and 1/4 of the $v$-hadrons are pseudoscalars (which, like pions, decay predominantly to heavy flavor).  The resulting branching to leptons is 9\%, hadrons account for 78\% (including $\sim 4.7\%$ to $\tau$), and missing energy from decays to neutrinos about 13\%.  The lifetimes are computed to be \cite{Strassler:2006im}:
\be
\Gamma_{\eta_v \rightarrow b \bar{b}} \sim 6 \times 10^{9} \mbox{sec}^{-1} \frac{f_{\eta_v}^2 m_{\eta_v}^5}{(20 \mbox{ GeV})^7} \left( \frac{10 \mbox{ TeV}}{m_{Z'}/g'}\right)^4
\label{eq:GammaEta}
\ee
\be
\Gamma_{\rho_v} \sim 4 \times 10^{13} \mbox{sec}^{-1} \frac{ m_{\rho_v}^5}{(20 \mbox{ GeV})^5} \left( \frac{10 \mbox{ TeV}}{m_{Z'}/g'}\right)^4.
\ee 
One can see that while the $\rho_v$ decays are typically prompt, the pseuod-scalar lifetimes may be long enough, for lower $v$-hadron masses, to give rise to a displaced vertex.  In this case, the displaced vertex is the preferred search method for Hidden Valleys, since the backgrounds could be rapidly eliminated.  As a result, we focus on the higher mass $v$-hadron case where the lifetimes are sufficiently short that no displaced vertex results, $i.e.$ for $m_{\eta_v} \gtrsim 30 \mbox{ GeV}$.

In our analyses below, we will decay the $v$-mesons to the SM leptons and quarks at the parton level.

\section{Separating Hidden Valleys From Standard Model Backgrounds}

In searching for new physics at hadron colliders, the most important aspect is to
identify the characteristic and distinctive features of the signal in order to distinguish 
it from standard model 
processes. Although a concrete prediction of the Hidden Valley signal will 
typically be model-dependent, there are general features that will guide us in the search. 
 Generically, 
 the signal events are less jetty, with higher sphericities, cluster masses, and typically multiple leptons in the final state, some of which may be isolated.  The presence of a low mass, narrow resonance in $\mu$ pairs from the $v$-hadron decay is the most distinctive signature in the model we consider.  All of these features can be used to reduce the standard model background significantly.  We turn to quantifying these features step by step.

As stated in the introduction, for definiteness, we will be studying $v$-hadrons with masses 
\begin{equation}
m_{vh} \gtrsim 30\ \mbox{ GeV}, 
\end{equation}
presumably determined by the confinement scale, while taking the $v$-quarks to be essentially massless. 
From Eq.~(\ref{eq:GammaEta}) we see that when $v$-hadrons are this heavy, no displaced vertex can typically be used to search for $v$-hadrons.  The mass of these hadrons in the absence of displaced vertices is typically much larger than standard model hadrons, which have masses less than the $B$ meson mass at 5 GeV.  
To this extent, we have chosen a rather challenging $v$-hadron mass for detection,
when there is no displaced vertex for $m_{vh} \gtrsim 30$ GeV.   Higher mass $v$-hadrons will be more easily distinguished from the soft QCD background of $b$, $c$ quark hadrons, as we argue in this section.

Most $v$-quark and $v$-antiquark pairs are produced, as we saw, at the $Z$ or $Z'$ resonance.  In what follows we separate signal on the $Z$ and $Z'$ peaks by the cuts Eqs.~(\ref{MZZcut}) and Eq.~(\ref{MZZpcut}), 
since the signals are qualitatively different there and they help with the background considerations.

Upon hadronization, an energetic $v$-quark gives a broad and massive $v$-jet on account of the higher hadronization scale, with a potentially high multiplicity of $v$-hadrons in it as already shown in Fig.~\ref{multiplicity}.  In constructing the events, we take a generous rapidity coverage and neglect
 particles with low transverse momenta. We thus adopt the acceptance for particle detection
 \be
 |\eta|<4.9, \quad  p_T^{} > 3\  {\GeV},
\label{eq:accept}
 \ee 
and ignore the charged particles outside this range. 
Each $v$-hadron may decay to a pair of muons with a branching fraction which is taken to be 4.7\%.  A pair of muons, even if they are fairly soft from the decay of the $v$-hadron, can be used as a trigger on the event. We
take the kinematical acceptance of the two muons as \cite{:1999fq}
\begin{equation}
p_T(\mu) > 10 \mbox{ GeV},\quad |\eta(\mu)|< 2.5.
\label{eq:mu}
\end{equation}
In Fig.~\ref{csdis}, the dashed curve gives the invariant mass distribution of the primary $v\bar{v}$ 
after the trigger cuts of Eq.~(\ref{eq:mu}).  We see a signal rate reduction of more than an order of magnitude
near the $Z$ peak, but only a modest reduction near the $Z'$. 
This trigger also helps considerably with the background separation. 
We will demonstrate the qualitative differences between the events from the signal
and the SM backgrounds in this section. In particular, for the sake of illustration, 
we will present the QCD background analyses for the processes 
\be
pp \to b\bar b X,\quad t\bar t X,
\ee
simulated by PYTHIA \cite{Sjostrand:2006za} with the full QCD initial state radiation, 
parton showering and hadronization.

 The $p_T$ distribution of the hardest lepton in the event for the Hidden Valley signal, 
 $b \bar{b}$ and $t \bar{t}$ events are shown in Fig.~\ref{fig:pTdis}.   In these figures and in what follows, the $b \bar{b}$ backgrounds are sufficiently large that the curves are multiplied by a factor ($10^{-3}$ in this case) so that they can be shown on the same scale as the signal.
 The presence of the large number of soft muons is mainly a result of the collinear 
 behavior of $b\bar b$ along the beam direction and the 
 softness of the muons from $B$ decays.
\FIGURE[t]{
\epsfxsize=2.9in \epsfbox{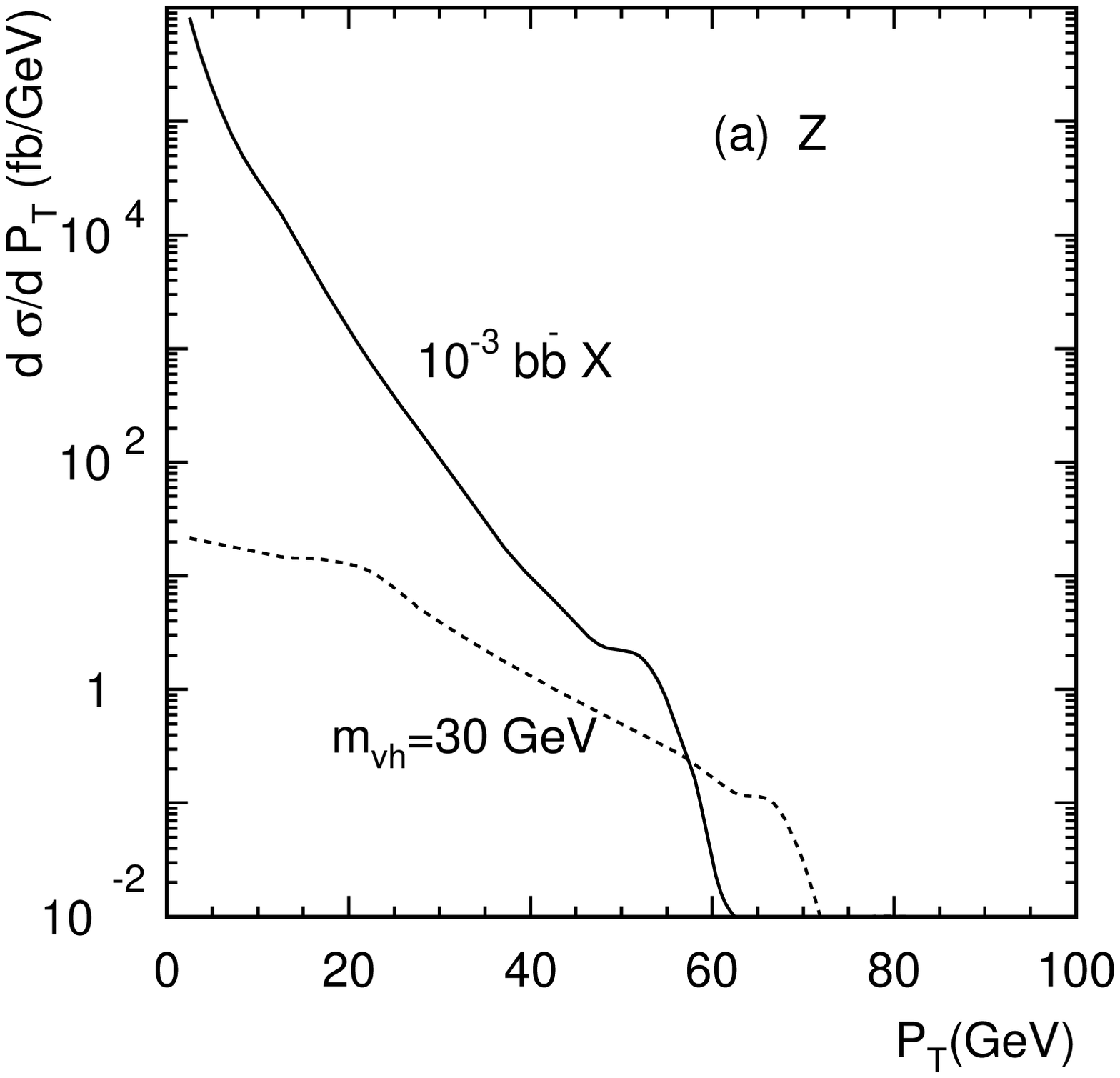}
\epsfxsize=2.9in \epsfbox{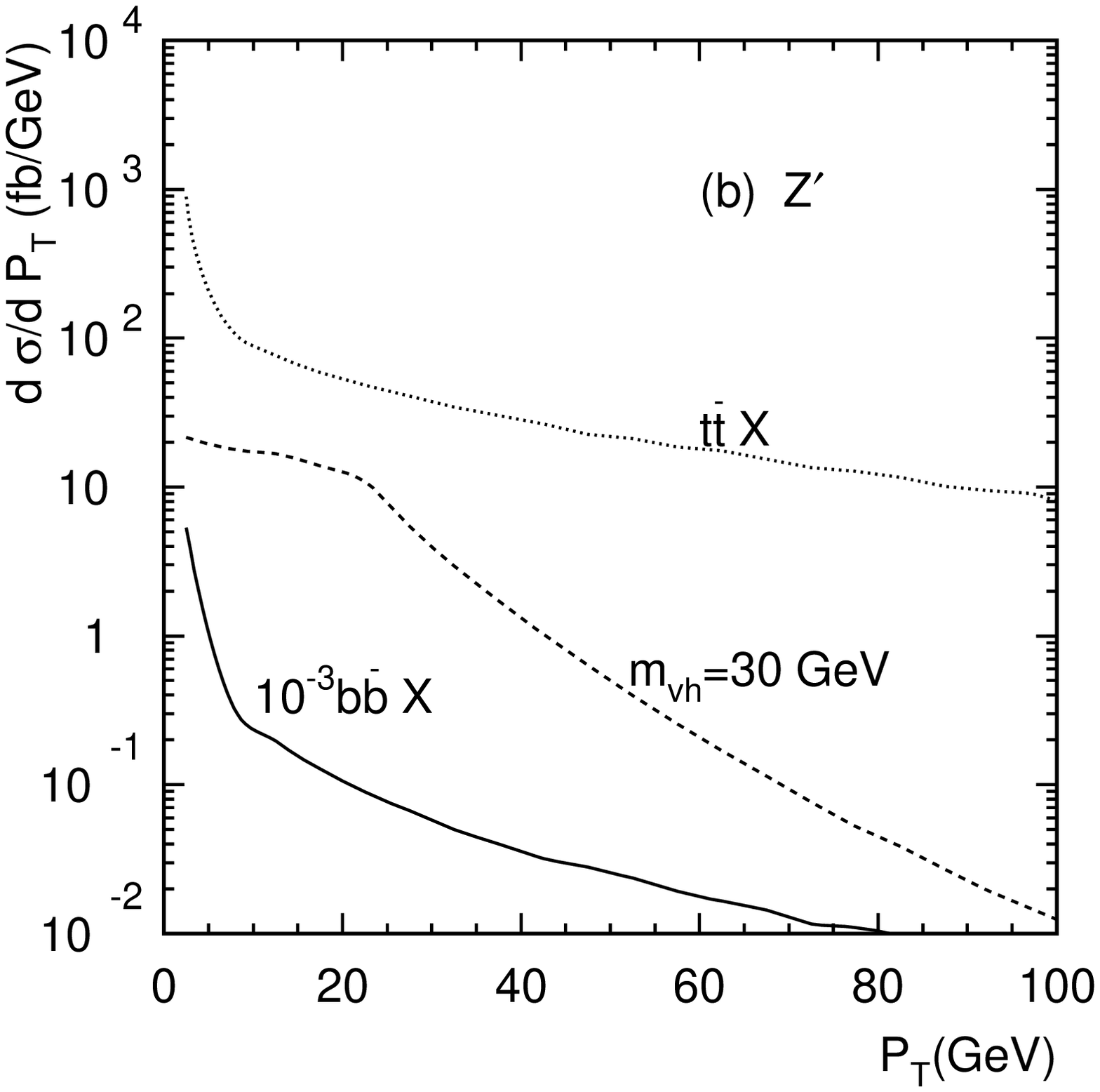}
\caption{$p_T$ distribution of the hardest lepton of Hidden Valley events and the QCD backgrounds, $b \bar{b}$ and $t \bar{t}$, after the particle acceptance, Eq.~(\ref{eq:accept}), has been imposed.},
\label{fig:pTdis}
}  

The pair of leptons from a $v$-hadron decay form a smaller angle than those from 
the SM Drell-Yan and $b\bar b,\ c\bar c$ type processes in which the leptons are typically back-to-back
in the transverse plane. 
We thus consider a cut for the relative angle  in the transverse plane between the two hardest leptons in the event (whether they be muons or electrons)
\be
\phi_{\ell\ell} < \pi/2 .
\label{eq:phi}
\ee
This cut will help eliminate most of the lepton pair events from the SM gauge bosons,
at very little cost to the signal. 

\FIGURE[t]{
\epsfxsize=2.9in \epsfbox{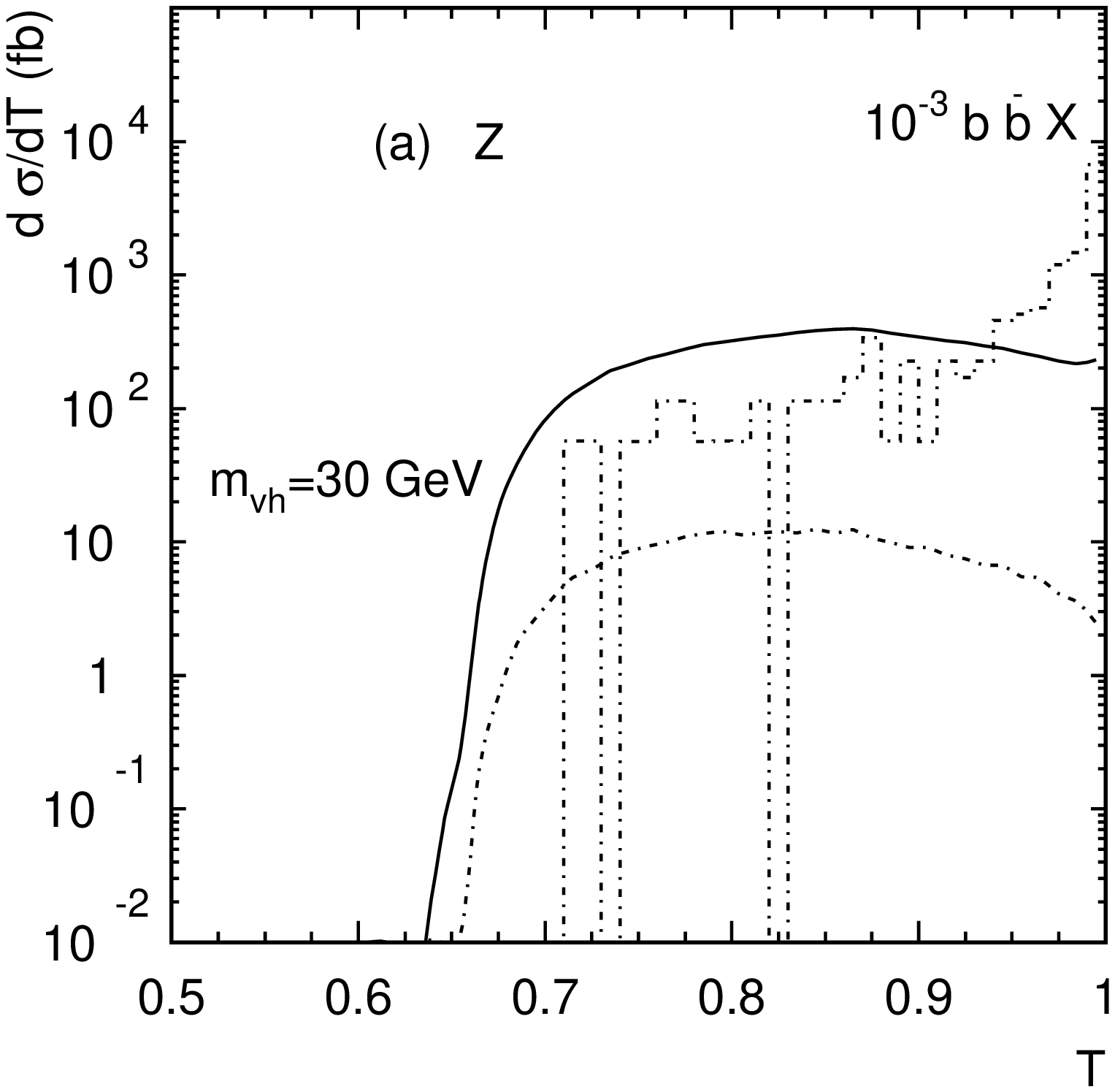}
\epsfxsize=2.9in \epsfbox{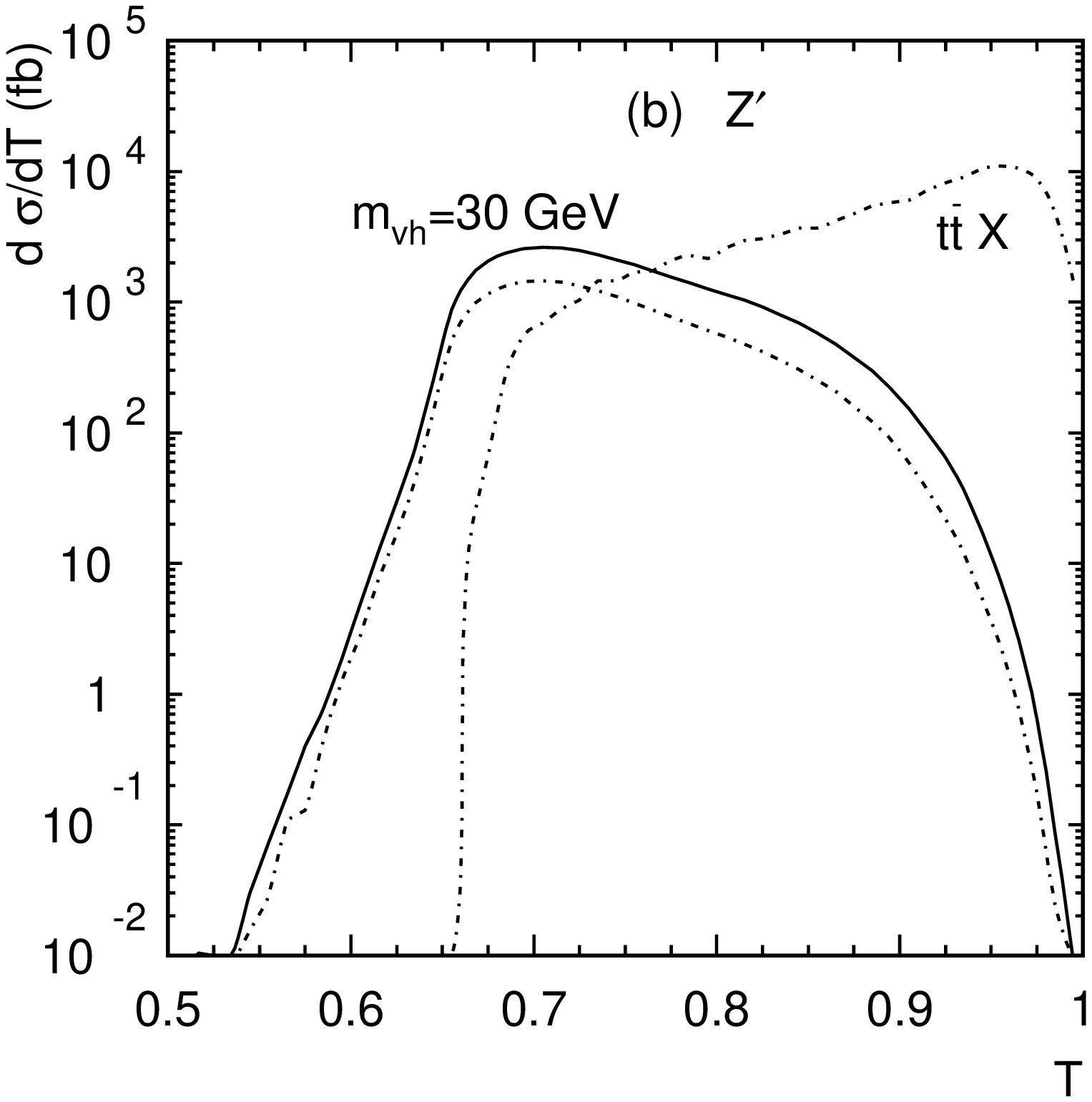}
\caption{Differential cross-section as a function of thrust $T$ for $m_{vh}=30$ GeV
and the backgrounds from $b \bar{b}$ and $t \bar{t}$ with (a) a $Z$ mediator and (b) a $Z'$ mediator. The solid lines indicate the distributions before the two muon trigger Eq.~(\ref{eq:mu}) and the cut on the angle between the leptons, Eq.~(\ref{eq:phi}), are imposed, 
while the dashed lines are after.}
\label{fig:T}
}

In order to demonstrate the effect of the high confinement scale and hadron mass on the shape of the event and on selected observables, such as the thrust, sphericity, and cluster mass, we compare in what follows the distributions for a 30 GeV $v$-hadron mass against dominant QCD backgrounds from $b \bar{b}$, $t \bar{t}$.
The collinearity of a jet is often quantified in terms of the thrust
\be 
 T \equiv \mbox{max}\left[ { \sum_i |{\bf p}_i \cdot {\bf n} | \over \sum_i |\bf{p}_i|} \right],
 \nonumber
\ee
where $\bf n$ is the thrust direction obtained my maximizing the sum.  
 We modify this expression in a way that is appropriate for an event in hadronic collisions, 
 namely to replace all momenta of the observed particles with their transverse components
 \be
 T = \mbox{max}\left[ { \sum_i |{\bf p}_{Ti}^{} \cdot {\bf n} | \over  \sum_i |{\bf p}_{Ti}^{} |} \right]. 
 \ee
 We compare the thrust distributions $d \sigma / d T$ in Fig.~\ref{fig:T}  for a  $v$-hadron signal
 and $b \bar{b}+X$ background centered on the $Z$ mediator (panel (a)) 
and $t \bar{t}$ background on the $Z'$ mediator (panel (b)).  We show the signal distributions both 
 before (solid curve)  and after (dashed curve) 
 the two muon cuts Eqs.~(\ref{eq:mu}) and (\ref{eq:phi}) are imposed.  
 The backgrounds are shown only after these cuts are imposed.
We can see that the $b\bar{b}$ background (and, by extension, the background from any low mass 
hadrons such as $D$ mesons) yields a more jet-like structure, with $T$ close to unity.
The cut
\be
T < 0.95\ {\rm for}\ Z,\quad T < 0.9\ {\rm for}\ Z'
\label{eq:tcut}
\ee
will very efficiently remove the backgrounds from $b\bar{b}$ and $c\bar{c}$ etc.  The long tail on the $b$ distribution in panel (a) is due to the effects of the initial state radiation.  The un-eveness in the curves results from a loss of statistics from the stringent muon cut, Eq.~(\ref{eq:mu}); this cut reduces an initial event sample of 10 million to just a few tens of events which are shown in the dashed $b\bar{b}X$ curve in the figures.
Alternatively, one can define the sphericity matrix in the transverse plane
\be
S(\alpha, \beta)\equiv {\sum_j p_j^{\alpha} p_j^{\beta} \over \sum_i |p_{Ti}^{} |^2} ,
\label{sphericity}
\ee 
with $\alpha,\beta = x,y$.  The two-dimensional sphericity  (also called circularity) is given in terms
of the two eigenvalues $Q_1$ and $Q_2$ by
\be
S = {2Q_1 \over Q_1 +Q_2} ,
\ee
with $Q_1 \le Q_2$ and $0\le S \le 1$.
We compare the distributions for a  $v$-hadron signal against the backgrounds from  $b \bar{b}$  and $t \bar{t}$
in Fig.~\ref{fig:sph}, again before and after the two muon cuts have been imposed.  We obtain some marginal improvement for both top and bottom backgrounds, where again the un-eveness in the bottom background curve is due to statistics.  We make a cut
\be
S > 0.1
\label{eq:scut}
\ee
for both $Z$ and $Z'$ mediators.

\FIGURE[t]{
\epsfxsize=2.9in \epsfbox{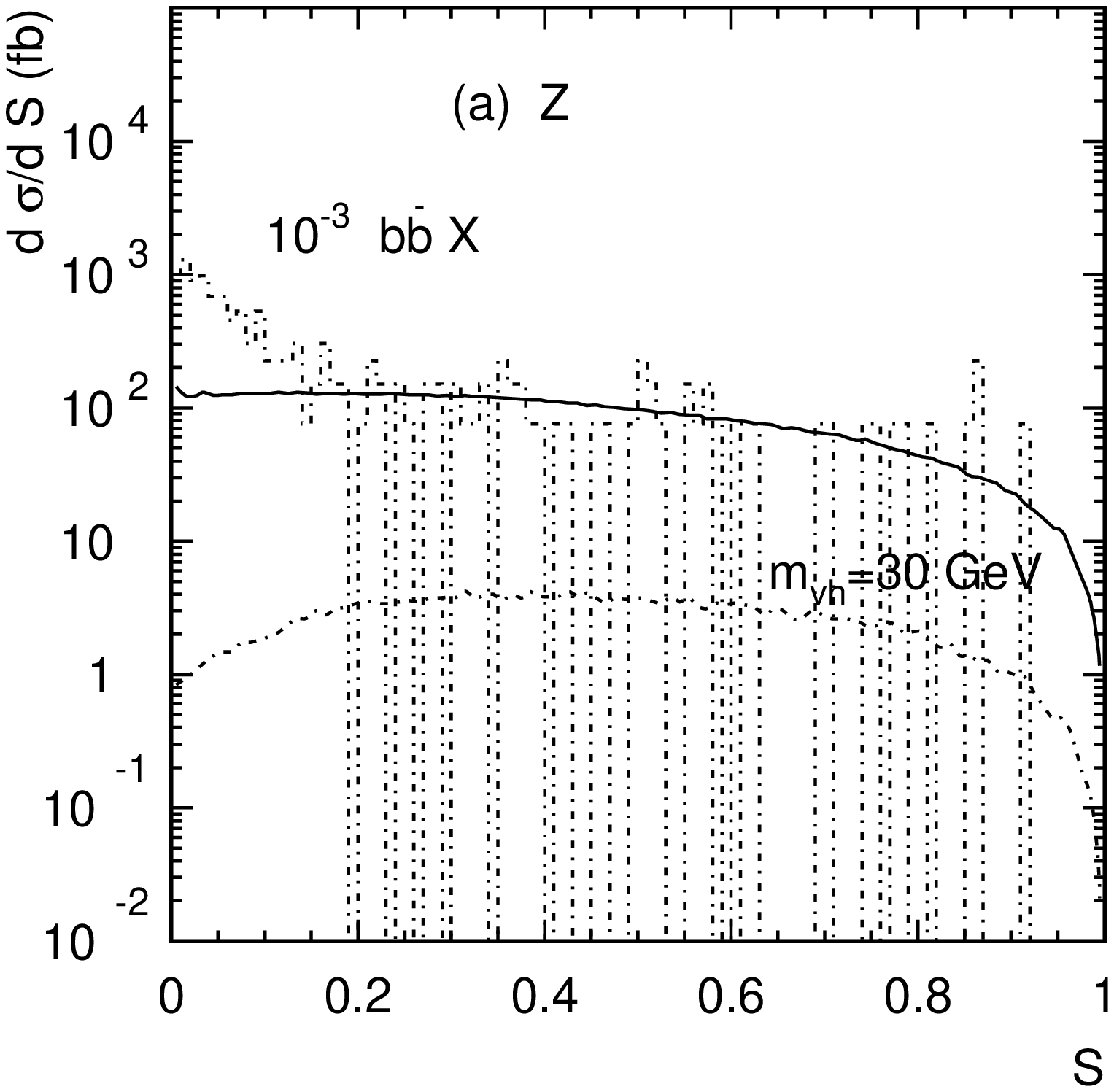}
\epsfxsize=2.9in \epsfbox{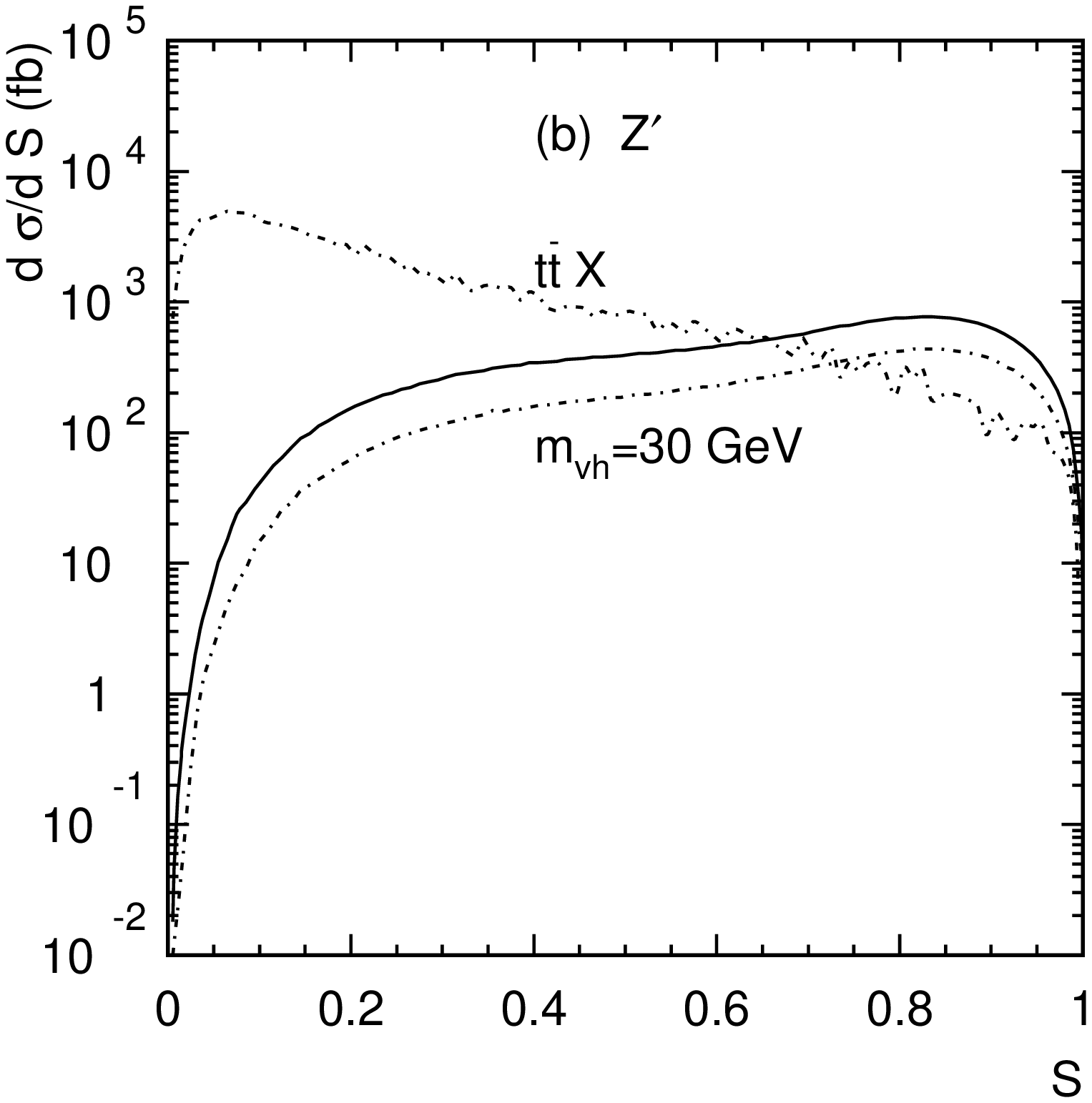}
\caption{Differential cross-section as a function of sphericity $S$ for $m_{vh}=30$ GeV
and the backgrounds from $b \bar{b}$ and $t \bar{t}$ with a (a) $Z$ mediator and (b) $Z'$ mediator before (solid) and after (dashed) the trigger, 
Eq.~(\ref{eq:mu}) and cut on the angle between the leptons, Eq.~(\ref{eq:phi}), are imposed. }
\label{fig:sph}
}


After obtaining the thrust axis in the transverse plane, 
we can define two clusters around the thrust directions, separated by a
longitudinal plane perpendicular to the transverse thrust axis.  
The invariant mass of the $v$-jet cluster will be substantially larger 
than that of a standard model jet.  We sum all observable particles in a cluster 
\be
M_{cluster}^2 =  \left( \sum_i E_i \right)^2 -\left( \sum_i {\bf p}_i \right)^2.
\ee
The cluster mass will typically be half the mediator mass for a fairly spherical event, becoming smaller as the event becomes increasingly jetty.  We show the cluster mass distributions in Fig.~\ref{clusterdis}
for $Z$ mediator (panel a) and $Z'$ mediator (panel b).  The $t \bar{t}$ and $b \bar{b}$ backgrounds after cuts of Eqs.~(\ref{eq:mu}) and (\ref{eq:phi}) are also shown. In contrast to a Hidden Valley event where the cluster mass 
is of the order of a half of the mediator mass, 
the cluster invariant mass for the heavy quarks should be peaked near $M_{cluster} \approx m_b,\ m_t$.
Thus we can further remove the heavy quark  backgrounds with a simple cut
\be
M_{cluster} > 
\left\{
\begin{array}{cc}
20 \mbox{ GeV} & {\rm at}\ Z,\\
200  \mbox{ GeV} & {\rm at}\ Z' .
\end{array}
\right.
\label{eq:mclustercut}
\ee
We would like to point out another important feature for the signal events.
The two clusters should have roughly equal cluster mass up to the fluctuations 
of the $v$ and $\bar v$ hadronization and the decay of the $v$ mesons. 
It can be seen from Fig.~\ref{clusterdis} that the mass spread is only 
about $30 \pm10$ GeV 
at the $Z$ pole and $250 \pm 50$ GeV at the $Z'$. 
For higher order backgrounds on the other hand, such
as $\mu^+\mu^-$+jets, $c\bar c,\ b\bar b$+jets, the event shapes would be rather
asymmetric on both sides. 

\FIGURE[t]{
\epsfxsize=2.9in \epsfbox{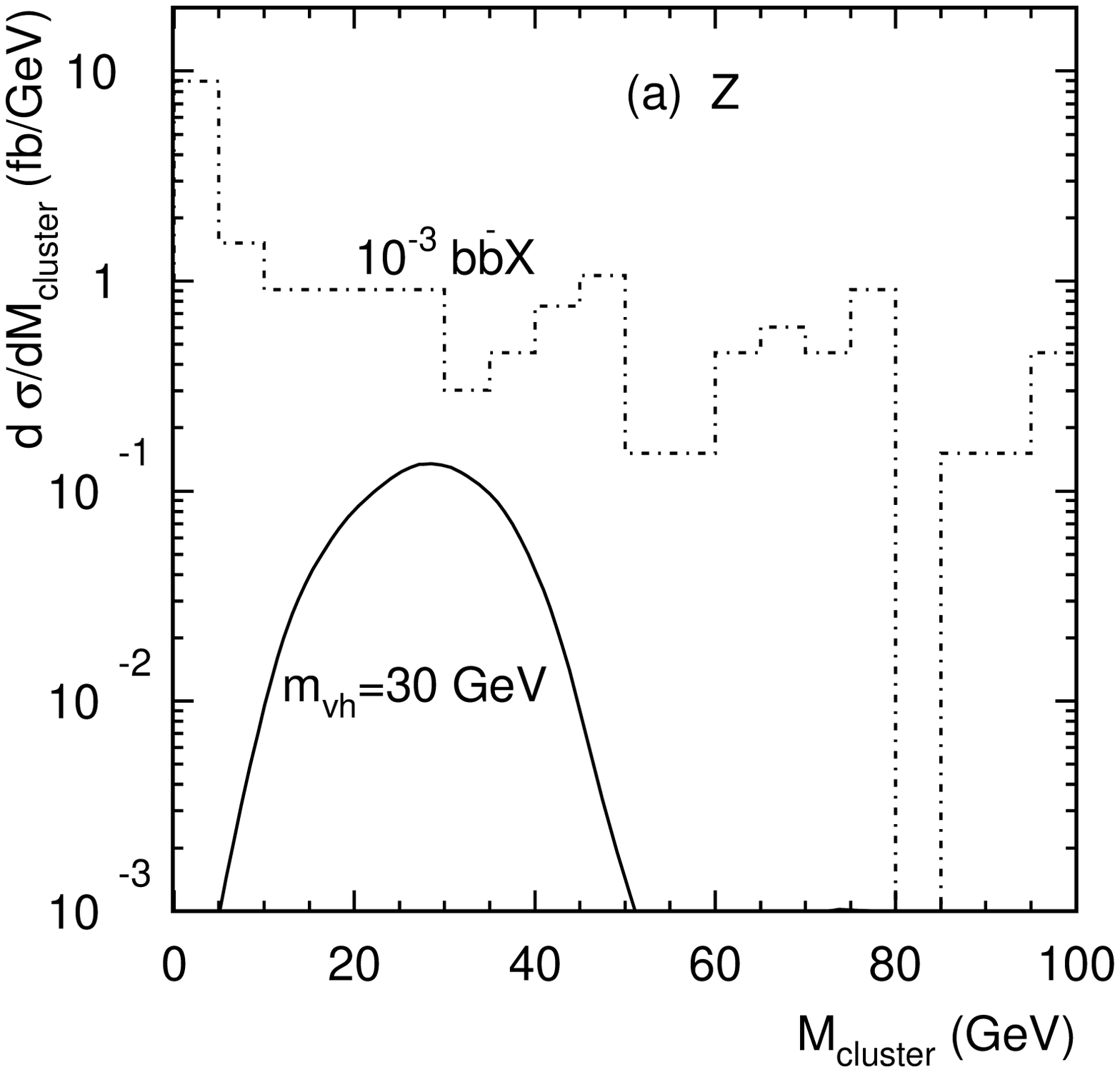} 
\epsfxsize=2.9in \epsfbox{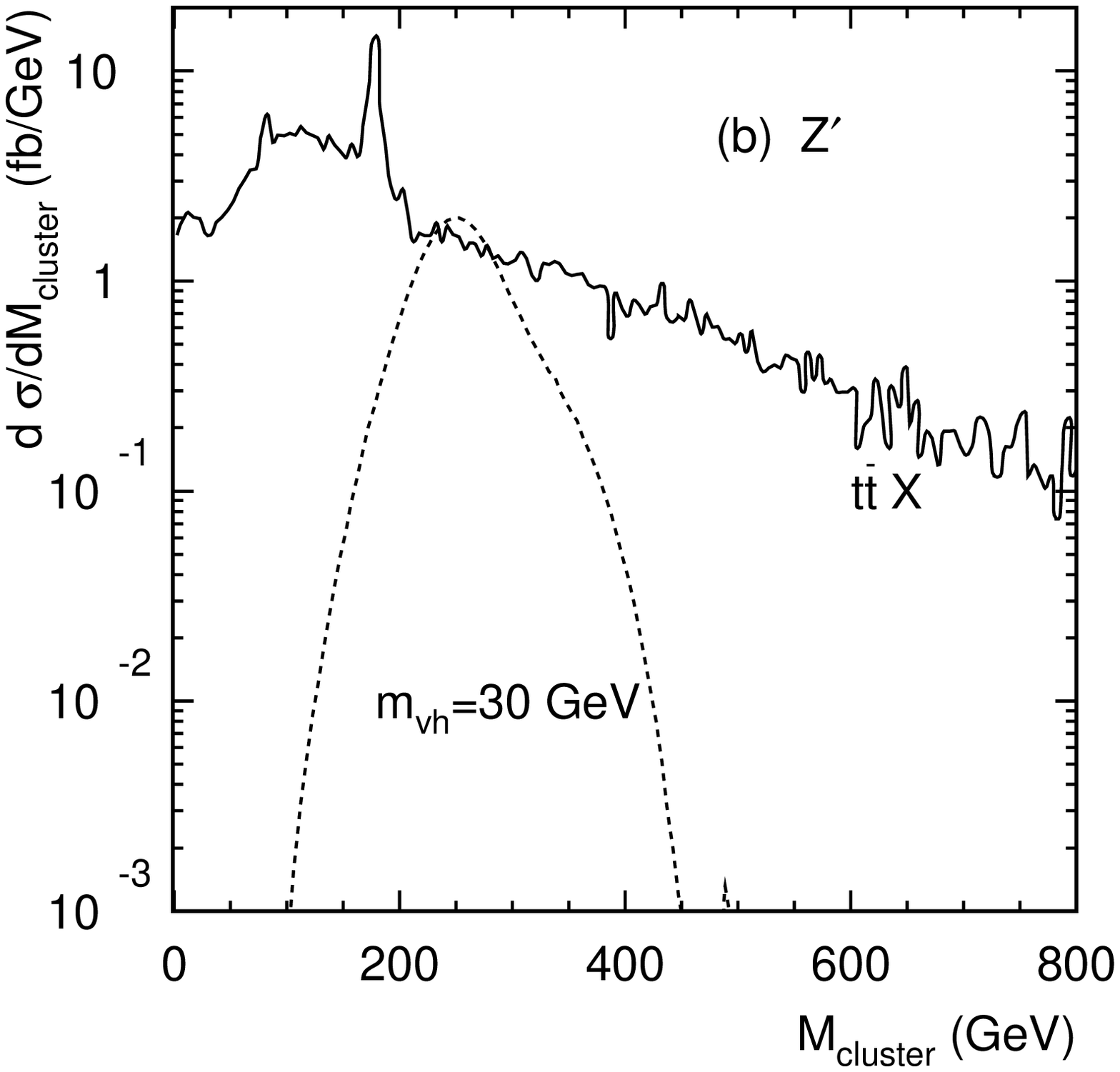}
\caption{Differential cross-sections  for 30 GeV $v$-hadron 
as a function of cluster invariant mass with (a) $Z$ mediator and (b) $Z'$ mediator.
%
The $b \bar{b}$ and $t \bar{t}$ backgrounds are also shown.  All curves are after cuts 
Eqs.~(\ref{eq:mu}) and (\ref{eq:phi}) have been imposed. }
\label{clusterdis}
}

In order to develop an intuitive picture of a typical  event satisfying the basic cuts, we show the 
 lego plots for two typical events satisfying the basic cuts Eqs.~(\ref{eq:mu}) and (\ref{eq:phi}) 
 in Fig.~\ref{events},  for one mediator each. 
 The height indicates the energy scale for hadronic energy (red), electromagnetic (blue),
 and muons (green). 
  As expected, these events are much fatter than QCD events, with high multiplicities, 
  and often many leptons in the final state.  The plots for the signal are at the lepton and quark level,
  without QCD hadronization. In a more realistic situation, the energy spread will be even broader. 
  To contrast Hidden Valley events with the backgrounds in the typical case, $b \bar{b}$, $t\bar{t}$ we consider here, we also show two typical events satisfying the same basic cuts in Fig.~\ref{backgroundevents}. We see clearly  the more isolated jetty structure. 
  The background events are generated by 
  PYTHIA \cite{Sjostrand:2006za} with the full QCD initial state radiation, 
parton showering and hadronization.

\FIGURE[t]{
\epsfxsize=2.9in\epsfbox{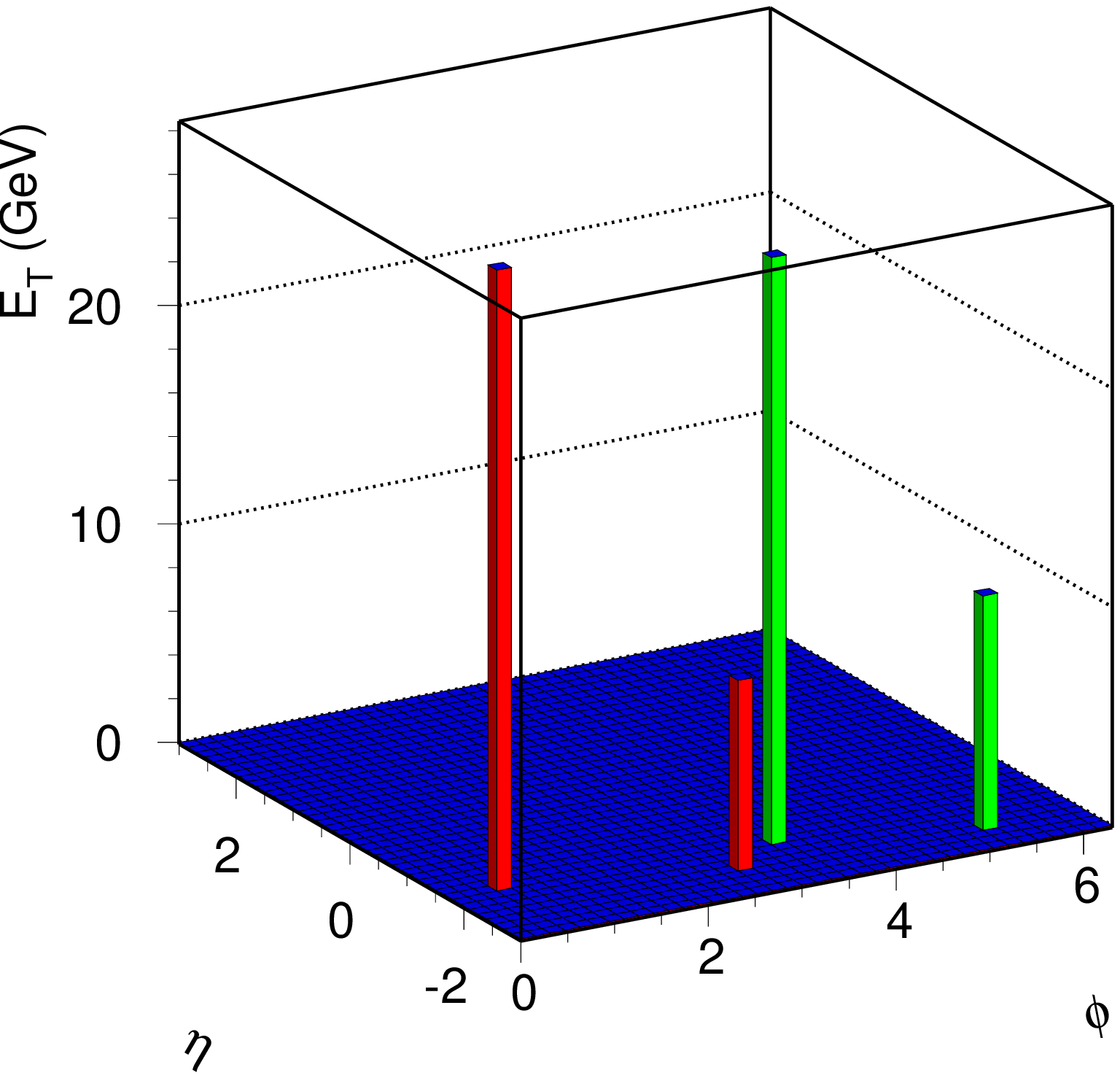}
\epsfxsize=2.9in \epsfbox{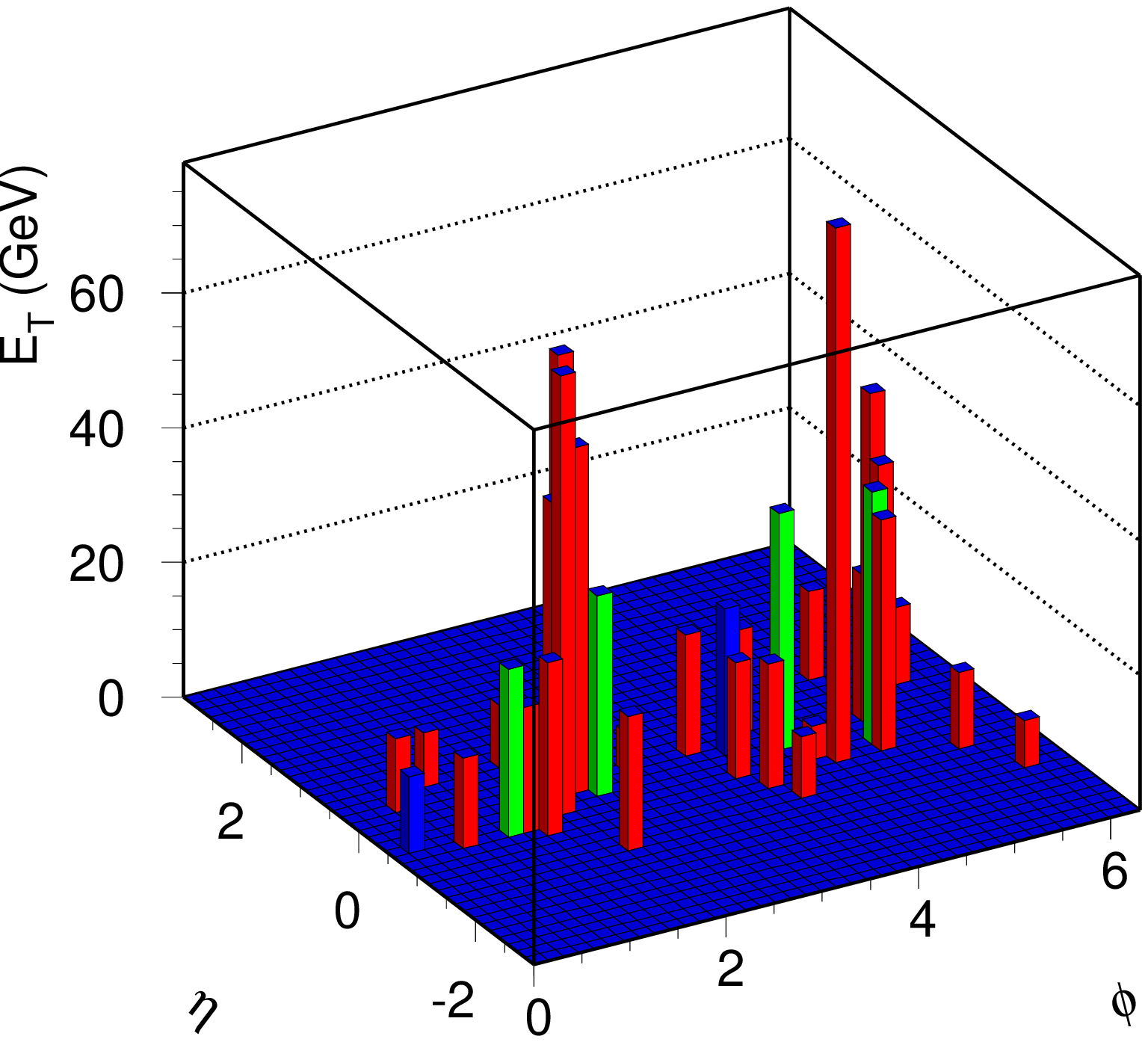}
\caption{Lego plots in the $\eta-\phi$ plane 
for typical events satisfying the basic cuts Eqs.~(\ref{eq:mu}) and (\ref{eq:phi}) for (a) a $Z$ 
mediator and (b) a $Z'$ mediator. 
The color codes indicate the hadronic energy  (red),
electromagnetic (blue), and muons (green).   Hadronic energy is at the parton level only (no showering).}
\label{events}
}
\FIGURE[t]{
\epsfxsize=2.9in\epsfbox{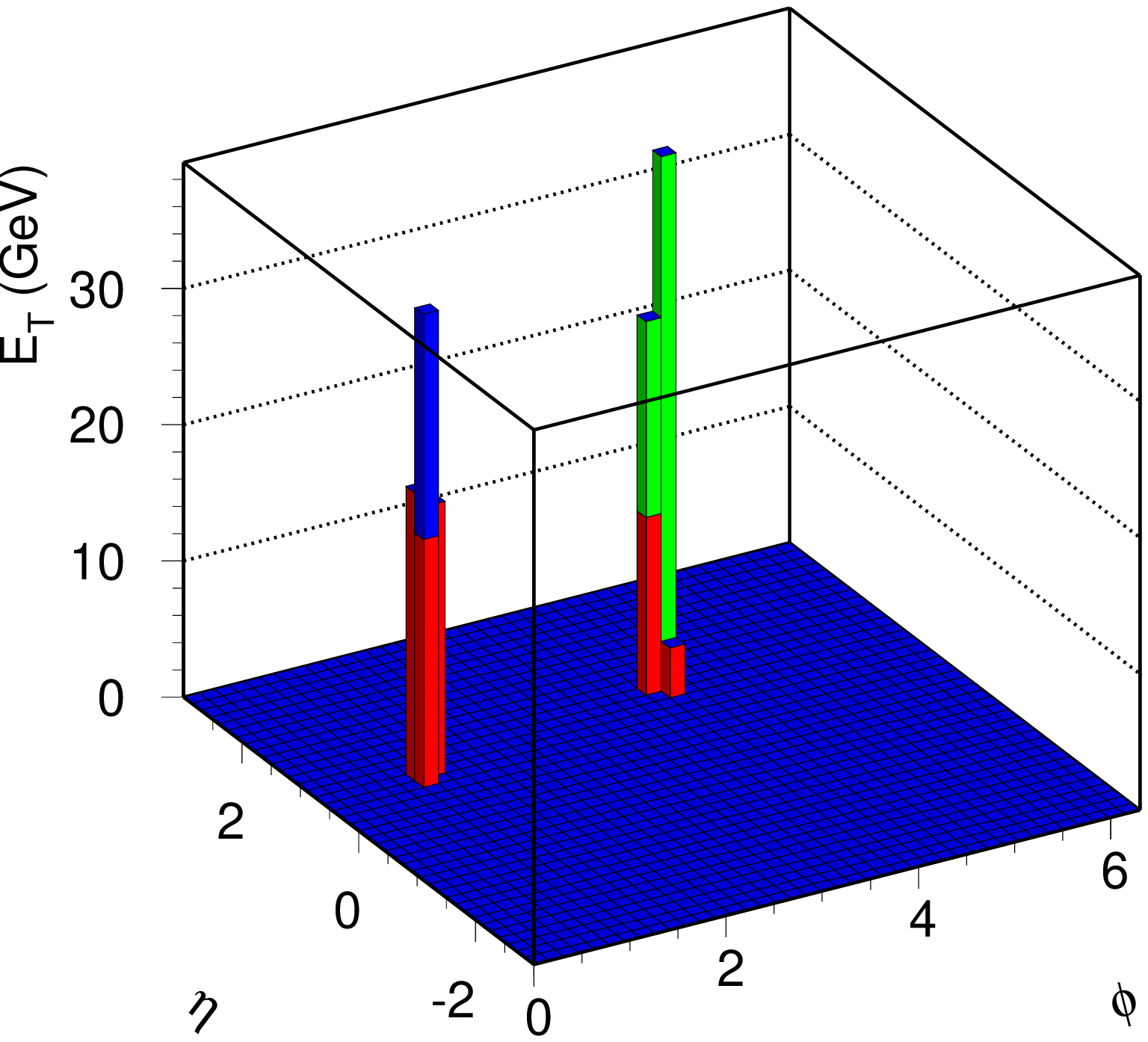}
\epsfxsize=2.9in \epsfbox{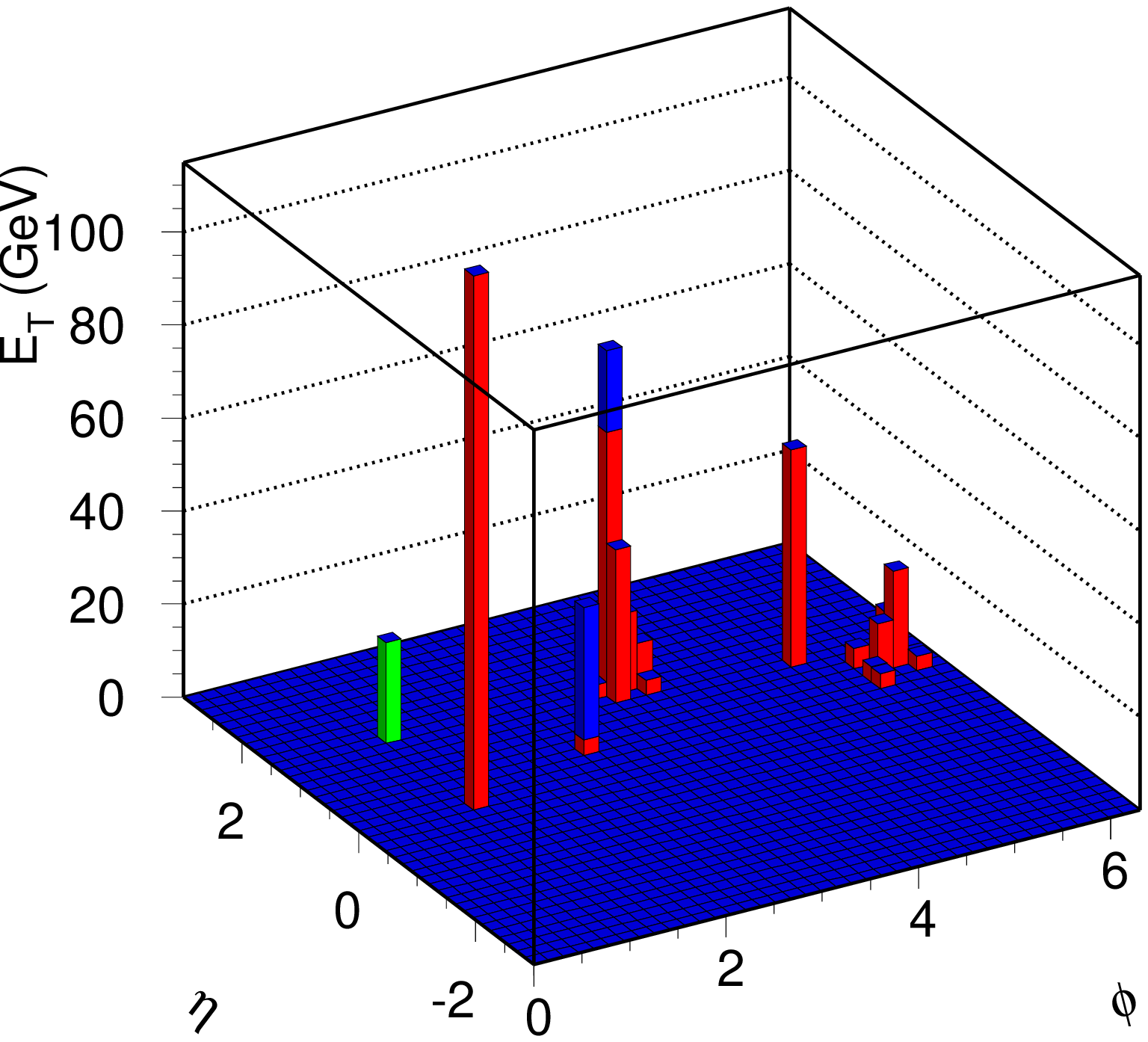}
\caption{Lego plots in the $\eta-\phi$ plane 
for typical (a) $b \bar{b}$ events (on the $Z$-mediator) and (b) $t \bar{t}$ events 
(on the $Z'$-mediator) satisfying the basic cuts Eqs.~(\ref{eq:mu}) and (\ref{eq:phi}).
The color codes indicate the hadronic energy  (red), electromagnetic (blue), and muons (green). }
\label{backgroundevents}
}

With this set of judicious cuts, we can now proceed to the decisive step, 
to reconstruct the $v$-hadron mass. 
This reconstruction can provide ultimate power to distinguish the signal from the backgrounds. 
We propose to reconstruct the $v$-hadron resonance from the leptonic decay products, in particular
a pair of muons. Quite often, there are more than 2 muons in an event 
($n_+\ \mu^+$'s and $n_-\ \mu^-$'s). 
We form all the  $n_+ n_-$ combinations of $\mu^+\mu^-$.  Among all possible opposite sign pairs in the event, if there are $k$ pairs with the same invariant mass ($\pm 1 \mbox{ GeV}$), we weight each pair with $1/k$, throwing the pairs away that do not match.  In this way we are able to select the correct muons to pair together.  In the case that we are missing a muon (from the detector acceptance) and none of the pairs match, we compute all possible pairs and weight it by $1/(n_+ n_-)$.  This is a combinatorial background that reduces the signal.
This is shown in Fig.~\ref{vhadronmass}, 
for both the $Z$ (panel (a)) and $Z'$ (panel (b)) mediators.  
With the excellent momentum resolution at these low muon energies, 
we take 500 MeV bins, consistent with experimental resolution at those energies \cite{:1999fq}, 
and show the number of events in each bin predicted for $100 \mbox{ fb}^{-1}$ (on $Z$ peak) and 
$10 \mbox{ fb}^{-1}$ (on $Z'$ peak) of data.  We also show the results from 
$b\bar b,\ t \bar{t}$ backgrounds in these two panels respectively, which demonstrate the final
effectiveness to separate the backgrounds. 
The solid lines (for both signal and background) are with the $2\mu$ trigger and $\phi_{ll}$ cuts only; 
the dashed lines (both signal and background) include all cuts in Table 2 through $M_{cluster}$.  
It is noted from Fig.~\ref{vhadronmass}(a) that the background muon pairs (solid and dashed)
are all from the $b$-quark cascade decays. This implies that the higher order contributions 
of high $p_T$ $b\bar b$ events may not have been adequately generated in the PYTHIA 
simulation. We thus perform a parton level calculation for $b\bar b$ plus a high $p_T$ jet
with $b\bar b\to \mu^+\mu^- X$. Its $m_{ll}$ distributions are shown in Fig.~\ref{vhadronmass}
for both $Z$ and $Z'$ mediators, with only the basic trigger implemented.  The shape variable cuts ($S$, $T$ and $M_{cluster}$) cannot
be properly implemented on this background since it is computed at the parton level, although such cuts may be very effective in reducing it. 
%
We can impose a final resonant mass  cut to estimate the signal significance
\be
 m_{vh} - 1 \mbox{ GeV} < m_{\ell \ell} < m_{vh} + 1 \mbox{ GeV},
\label{eq:mll}
\ee 
where $m_{ll}$ is the invariant mass of the lepton pair selected for the event.  Often there is 
more than one muon resonance in the event which can even further strengthen the signal
identification. 

We have now finished laying out the cuts which will distinguish the Hidden Valley Model from the Standard Model events.  We summarize these cuts, laid out step-by-step in 
Eqs.~(\ref{eq:mu}),~(\ref{eq:phi}),~(\ref{eq:tcut}),~(\ref{eq:scut}),~(\ref{eq:mclustercut}),~(\ref{eq:mll}), 
in Table~\ref{summary}, where we show their effect on the signal rate. 
The entries in the last  row show the $t\bar t$ background with the consecutive cuts 
corresponding to the signal of a $Z'$ mediator. This background
becomes much smaller than the expected signal. 
We have not included the backgrounds for the case of a $Z$ mediator, since 
there are subleading, but possibly significant processes like $b \bar b +$ jets that have not been adequately
simulated in PYTHIA with showering. Further work is required to show quantitatively and conclusively that the signal can be dug out of the background on the $Z$ peak.  By contrast, we have shown that the signal should be easily separable from background on the $Z'$ peak.

\FIGURE[t]{
\epsfxsize=2.8in \epsfbox{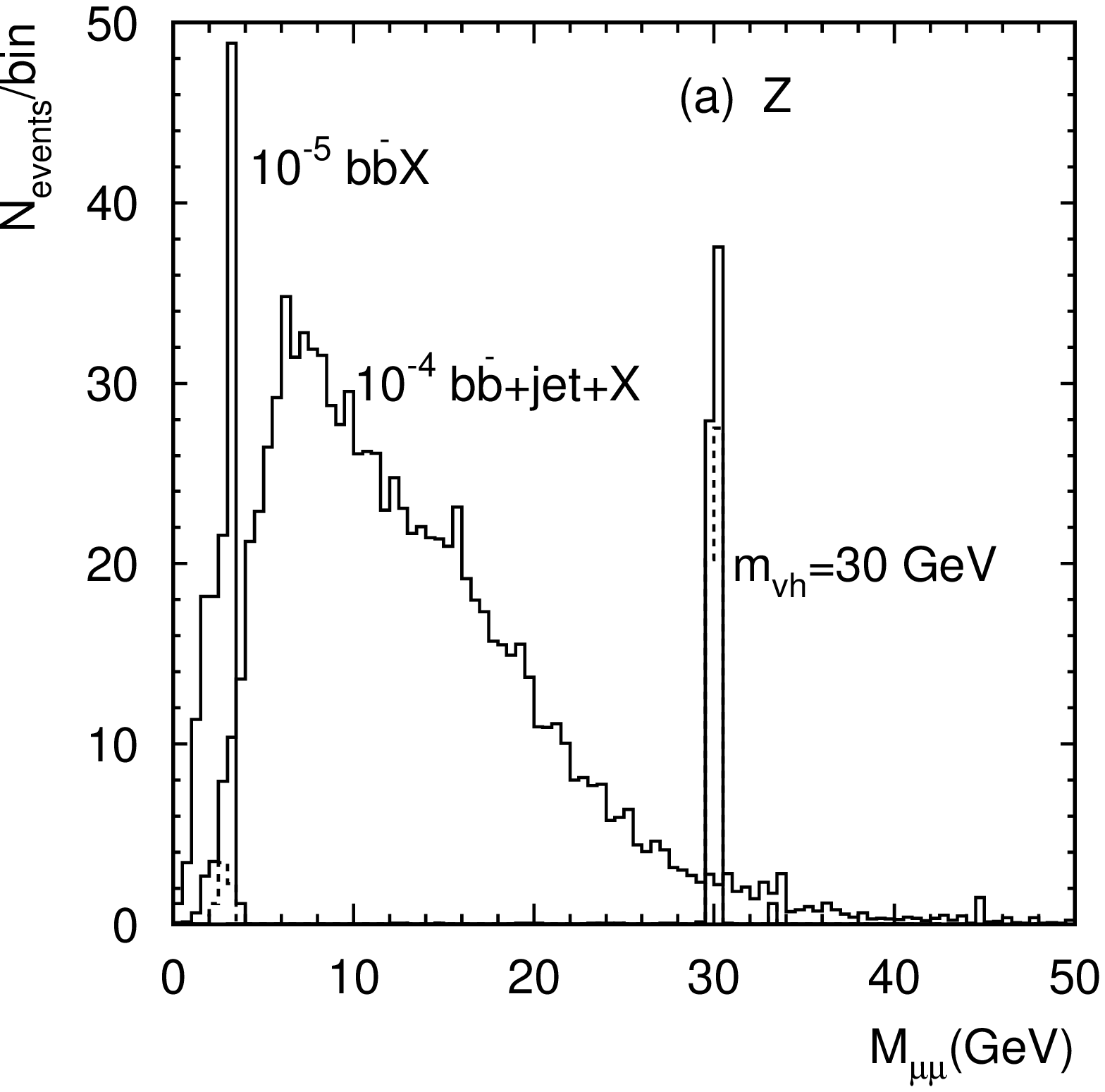}
\epsfxsize=2.8in \epsfbox{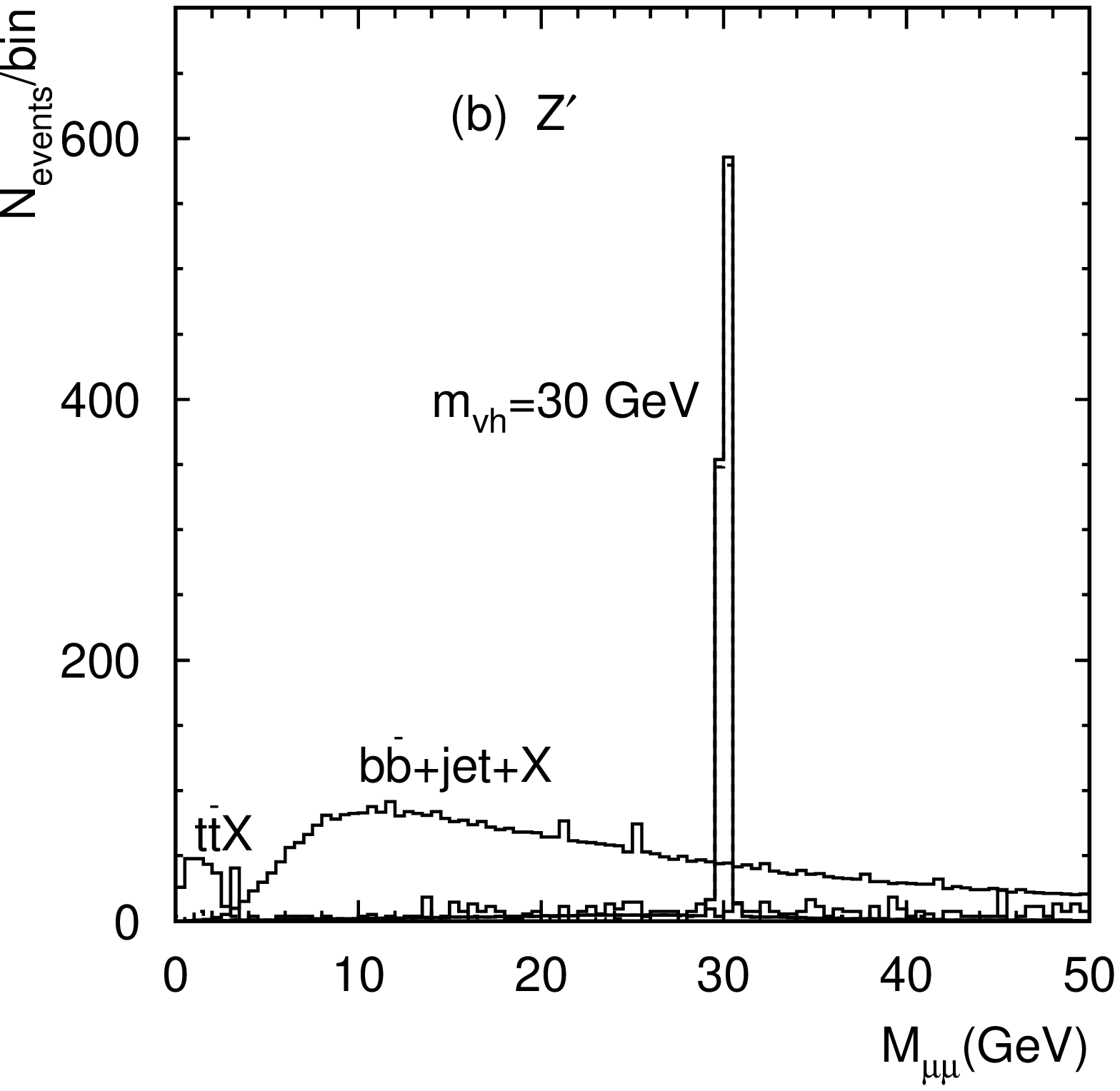}
\caption{The mass reconstruction of the $v$-hadron via the lepton pairs in the decay,
(a) for the $Z$ mediator and (b) for the $Z'$ mediator.   The bin size is 500 MeV, consistent with experimental resolution at those energies. 
The solid lines (for both signal and background) are with the $2\mu$ trigger and 
$\phi_{ll}$ cuts only; the dashed lines (both signal and background) 
include all cuts in Table 2 through $M_{cluster}$. The background curves for 
$b\bar b$+jets are before the cuts on $S,T$ are imposed.}
\label{vhadronmass}
}

The cuts that we have specified to this point show the main features of the Hidden Valley signal which will separate it from background.  There is an additional feature, however, which may also be useful for the signal separation should it be necessary.  When the $v$-hadrons decay to pairs of standard model particles via the mediators, 
the separation between the decay products (a pair of leptons or a pair of quarks) will be larger than is typical of standard model hadron decays, since the angular size of a typical decay is $m_{vh}/2E_h$, where $E_h$ is the energy of the $v$-hadron.
As a result of the more spherical jet structure of Hidden Valley events and the larger separation between decay products, an isolated lepton can be a feature to select for Hidden Valley events.  This is demonstrated in Fig.~\ref{dRmin}, where $\Delta R_{min}^{isol}$ is the separation between the most isolated lepton (whether muon or electron) and its nearest (non-leptonic) neighbor.  An additional cut on the leptons may then be designed so that
\begin{equation}
p_T(\ell) > 6 \mbox{ GeV},\quad |\eta(\ell)|< 2.5, \quad \Delta R_{min}^{isol} > 0.3.
\label{eq:muisol}
\end{equation}
This will substantially remove the SM backgrounds, especially those from heavy quarks,
while necessarily reduce the signal rate as well, depending on the number of isolated
leptons required, as can be seen from 
the last three columns in Table~\ref{summary} 
before imposing the $m_{vh}$ cut. 

%
%
\FIGURE[t]{
\epsfxsize=2.9in\epsfbox{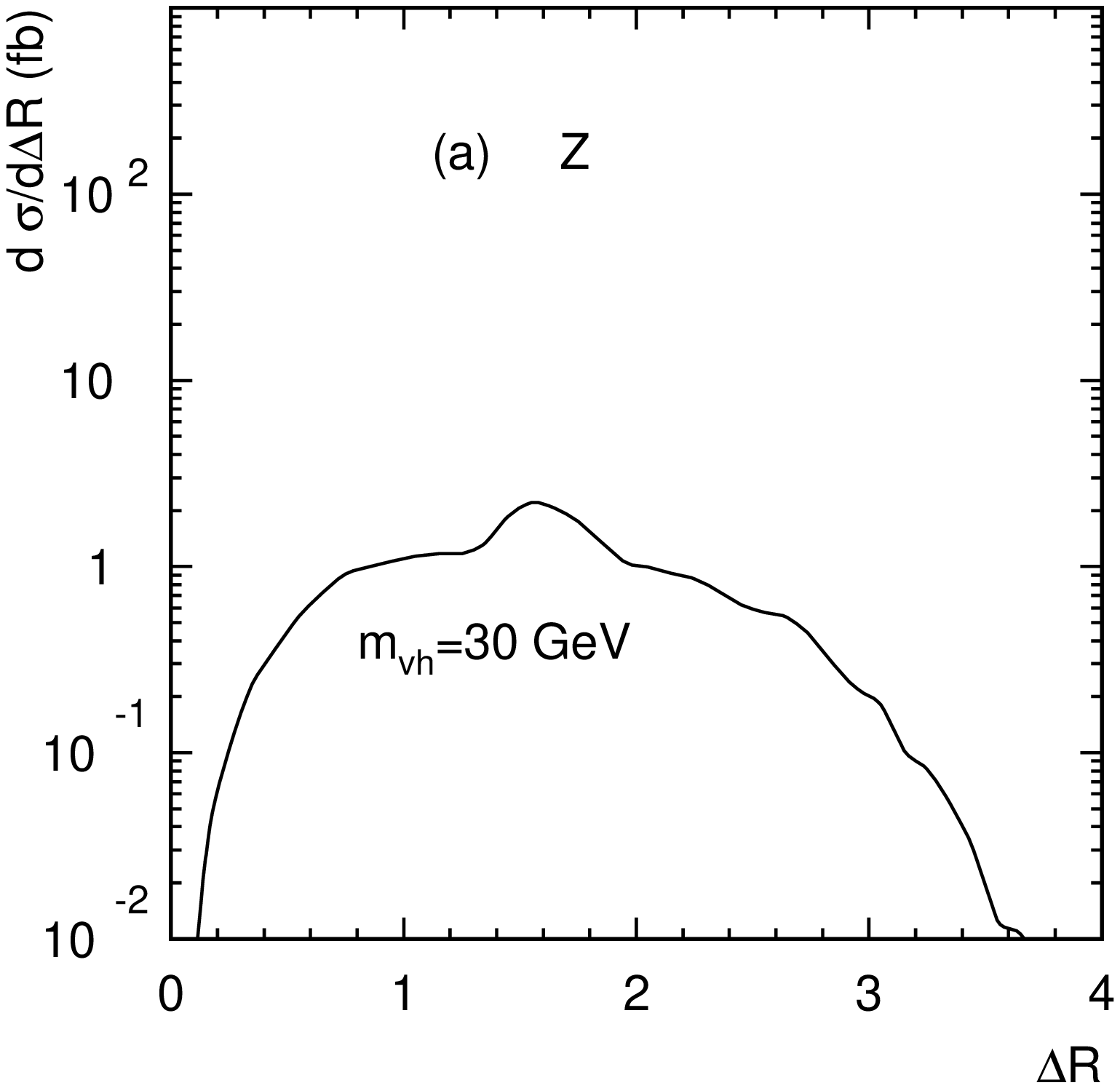}
\epsfxsize=2.9in\epsfbox{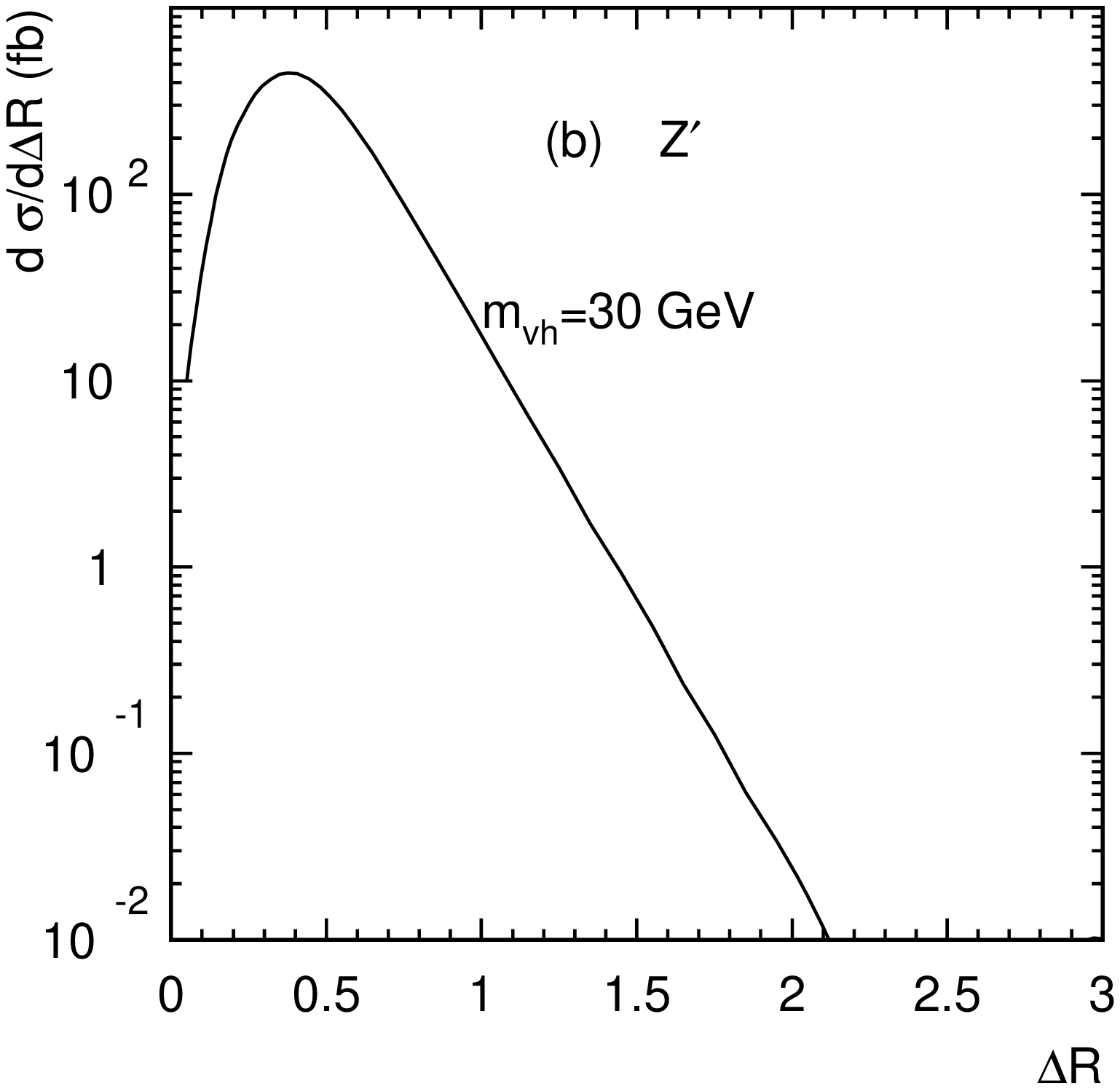}
\caption{Differential cross-section as a function of the separation for the 
 most isolated lepton $\Delta R_{min}^{isol}$ for $m_{vh}=30$ GeV  with (a) a $Z$ mediator and (b) a $Z'$ mediator.  }
\label{dRmin}
}

\TABLE[t]{
\begin{tabular}{|c|c|c|c|c|c||c|c|c|} \hline
& no & Trigger:    &  Thrust \&           &        &     &
\multicolumn{3}{c|}{isolated $\ell$'s}    \\  \cline{7-9}
& cuts &  $2 \mu's+\phi_{\ell\ell}$ & Sphericity &  $M_{cluster}$  &
$m_{vh}$ &    \multicolumn{3}{c|}{Eq.~(\ref{eq:muisol})}    \\ \hline
&fb& Eqs.(\ref{eq:mu},\ \ref{eq:phi}) & Eqs.(\ref{eq:tcut},\ \ref{eq:scut}) & 
Eq.(\ref{eq:mclustercut})& Eq.(\ref{eq:mll}) & 1 & 2 & $3 \ell$ \\ 
\hline
$Z$  & 125 & 0.69  & 0.62 & 0.50   & 0.46 & 0.50  & 0.13 & 0.011 \\
\hline
$Z'$        & 368  & 122 & 121 & 121  & 95  & 101 & 39  & 7  \\ 
$t\bar{t}X$ & 7960 & 207 & 79  & 36   & 0.4 &  28    & 23    &
0.2 \\
\hline\hline
\end{tabular}
\caption{Signal cross-sections (in fb) near the peaks of the $Z$ and $Z'$ mediators
with various kinematical cuts (defined in the marked equations in the text).
Cuts up to $m_{vh}$ are implemented cumulatively, while the cuts to the right of the double line
are implemented independently of each other, but in tandem with the basic cuts
(those cuts through $M_{cluster}$).  We also show the effect of the cuts on the dominant 
background  $t \bar{t} X$ for $Z'$ mediator.}
\label{summary}
}

We hope that by now we have demonstrated enough characteristics for the Hidden Valley
signal and have developed the judicious cuts to separate it from the SM backgrounds.  We have treated in detail the backgrounds which are the most dangerous.
We summarize now the potential backgrounds, 
with reference to how the cuts we have defined in this section have helped to remove them.  
Of course, all of these backgrounds are continuum in $m_{\mu\mu}$ distribution, 
above which a new narrow resonance may easily stand out.
\begin{itemize}
\item Drell-Yan background, that is, $Z$ and $\gamma^* \rightarrow \mu^+ \mu^-$ plus jets, is of order 1 nb, and is reduced on the $Z'$ by the requirement of (\ref{MZZpcut}) to the order of 7 fb.  To obtain two hemispheres of large cluster mass requires of order three additional hard jets, reducing the background below problematic levels.  $W$ plus heavy flavor can similarly produce dileptons, but is similarly reduced.

\item Gauge-boson pair background:  Another source of di-muon events may come from the gauge
boson pair production $W^+W^-, WZ, ZZ \to \mu^+\mu^- + X$. The production rates for these
processes start at the order of 100 pb, and are reduced to the order of 500 fb on the $Z'$.  Dimuon branching fractions reduce this by almost two orders of magnitude, and to obtain high cluster mass on both sides requires additional jets.

\item $c\bar c,\ b\bar b$ backgrounds:  A more difficult background results from the QCD strong production of heavy quarks $c\bar c$ and $b\bar b$ that can
lead to a large cross section of the order 1 $\mu$b at $p_T\approx 30$ GeV. 
With their semi-leptonic decays to $\mu^+\mu^-$ with approximately $10\%$ branching fraction and 
the basic triggering cuts, the rate will again be at a level of 1 nb.
Similar considerations
as those to eliminate the DY background, along with an additional requirement that
the system mass is near either the $Z$ or $Z'$ mediator, two massive clusters, 
and anti-tagging for $c,\ b$ will reduce this background by many orders of magnitude. 
We have shown by PYTHIA simulations that the leading order in $\alpha_s$ $b\bar b$ background process
could be effectively separated from the signal, as seen in Fig.~(\ref{vhadronmass}). 

Additional radiation of jets or multiple heavy quarks may make the event topology 
more spherical. However, the qualitative features, such as the jetty structure, non-equal cluster
masses, lepton isolation etc., are expected to provide additional handles to
discriminate those higher order backgrounds. For illustration, 
we calculated the process of $b\bar b$ plus a high $p_T$ jet
with $b\bar b\to \mu^+\mu^- X$ in Fig.~(\ref{vhadronmass}). This background is under control at the $Z'$ peak,
but sizable at the $Z$ peak. Nevertheless, we expect that it may be sufficiently
suppressed  by the sphericity, thrust, cluster mass, and in particular the lepton isolation, though simulations beyond the scope of this paper would be required to confirm this.
\item $t\bar t$ background: The top-quark pair production at the LHC has a total cross 
section of 800 pb, and is reduced on the $Z'$ to about 20 pb. Di-muon decays of the top quarks cost another factor of 80, but typically the angle between the two muons is very large, since the top quarks are boosted.  The most likely way to obtain two muons passing our cuts is from $t \rightarrow Wb$ in which both $W$ and $b$ produce a muon; since this can happen for either $t$ or $\bar{t}$ it costs a factor of about 40.  With the cluster mass requirement, our study has shown that the dimuon invariant-mass continuum background is small.  Additional jets can push the event above the cluster mass requirements, but the signal is so far above background that it will still easily stand out.  

\end{itemize}
Since we have already seen that our signal stands out far above the $t \bar{t}$ background, we are confident that inclusion of a larger set of continuum backgrounds and of detector backgrounds will not change our basic conclusion.  If it does, then a lepton isolation cut will eliminate the largest backgrounds, without costing too much of the signal, as is clear from Table 2.

We summarize the total cross-section for the signal in Table~\ref{summary} including all the cuts
and leptonic selections step by step. Based on the above arguments on the background rates
and their possible suppressions, we are led to believe that  the SM backgrounds are effectively 
removed.  The signal rate at the $Z'$ resonance is higher than that at the $Z$ by two orders of
magnitude with out parameter choices of Eq.~(\ref{para}). 
We can thus estimate the sensitivity for the signal observation for different values of
the parameters. Assuming an integrated luminosity of 10 fb$^{-1}$ and take 4 background
events as in Table \ref{summary}, this requires 15 signal events to reach 
a $99\%$ CL observation with Poisson statistics. We can thus reach a coupling strength
\be
g' \sim 1.7\times 10^{-2}\quad {\rm for}\  m_{Z'}=1\ {\rm TeV}. 
\label{reach1}
\ee
Conversely, if we fix the coupling, we obtain a $99\%$ CL sensitivity on the mass 
\be
m_{Z'}\approx 4.6\ {\rm TeV},\quad {\rm for}\ g'\sim 1/7,
\label{reach2}
\ee
after including the fall of the valence quark luminosity at high momentum fraction $x$ values, assumed to go like $(1-x_1)^4(1-x_2)^4$.

We reiterate that we have not performed detailed simulations including the
realistic experimental environment and detector effects, which would be necessary
to draw a more definitive conclusion.

\section{Summary and Conclusions}

In this paper, we have studied the feasibility of detecting hidden sectors with low mass bound-states
at hadron colliders. We assumed a generic gauge boson $Z'$ at the TeV scale as the mediator 
(that also mixes with the SM $Z$) to induce the interactions with the SM sector.  Most of the Hidden Valley signal results from a $Z$ or $Z'$ produced on-peak which then decays to Hidden Valley $v$-quarks. The $v$-quarks subsequently confine into $v$-hadrons, which then decay to SM pairs of quarks or leptons.
We chose the characteristic case with a pair of muons in its decay. 
We took the confinement scale to be $m_{vh} \gtrsim  $ 30 GeV, below which the $v$-hadrons
may be long-lived and thus would lead to different signatures. We demonstrated the characteristic
features of the signal with a Monte Carlo simulation of the hadronization, 
and contrasted them with the SM background expectations.
We showed that  accounting for the constraint from the $Z$ pole physics at LEP,  
searches at Tevatron may be difficult in this particular class of Hidden Valleys.
We thus concentrated on the searches at the LHC. 
We found that for the signal events
\begin{itemize}
\item there may be multiple hard muons ($p_T > 10 \mbox{ GeV}$), more widely separated, which can be used as triggers,
as seen in Figs.~\ref{fig:pTdis} and \ref{dRmin}; 
\item the events are more spherical and less thrusty, 
 as shown in Figs.~\ref{fig:T}, \ref{fig:sph} and \ref{events};
\item the invariant mass of the $v$-jet is high: $v$-jets are very ``fat'', 
and a signal event has a pair of back-to-back clusters of nearly equal mass,
as shown in Fig.~\ref{clusterdis};
\item muon pairs from the spin-one $v$-meson decay can reconstruct $m_{vh}$ as a narrow resonance, and provide the most effective kinematical variable to identify the new signal, as shown in Fig.~\ref{vhadronmass}.  Often there is more than one such resonance per event.
\end{itemize}
 The signal rate near $Z'$ is substantially larger
than that at $Z$, especially after imposing acceptance cuts on the leptons.   We have designed judicious cuts to remove or reduce the SM background, the most difficult of which are 
$b\bar b$ at high $p_T$, but relatively low center of mass energy (near the $Z$ peak).  The signal on the $Z$ peak is much more challenging than on the $Z'$ peak, due to the larger $b$ quark backgrounds.  By contrast, we have shown that the $t \bar t $ backgrounds on the $Z'$ mediator can be easily and efficiently removed.
%
%
On the $Z'$, we found large coverage for the Hidden Valley parameter space ($g',\ m_{Z'}$) at the LHC as shown
in Eqs.~(\ref{reach1}) and (\ref{reach2}).

We emphasize that although we studied a particular class of Hidden Valley models, the implications of our efforts are much more general.  Light spin-one resonances that can decay to dilepton pairs, and that are produced only in rare decays of the $Z$ or in decays of a new particle such as a Higgs or $Z'$, may easily arise in a very wide array of Hidden Valley models.  They may also arise in yet other classes of models.  A search for such resonances cannot succeed in a fully inclusive dimuon channel.  Instead, one must select events on the basis of variables such as total visible energy and event-shape observables; other examples might include missing energy, numbers of $b$ tags, etc.  These variables cut away most standard model processes, and with the events that remain, narrow dilepton resonances stand out easily, even for signals with small cross-sections.  We view this as a general lesson for LHC, and even Tevatron, search strategies.

We have studied the basic features of models with light quarks, confinement scale above $\sim 10 \mbox{ GeV}$, a light spin-one resonance, and a gauge mediator.  We note that there are other Hidden Valleys which remain to be explored.  
The phenomenology
depends on the matter content of the hidden sector, the size of the mass gap (confinement scale), and the nature of the mediator, each of which may have unique implications for general collider searches for Hidden Valleys, as well as for astrophysics and cosmology.  
With these tools to study low mass hidden sectors at hadron colliders, we may uncover a richer sector beyond the mostly discussed new physics at the
TeV scale.

\acknowledgments{This work is supported 
in part by the US Department of Energy, under grant DE-FG02-95ER40896,
 the Wisconsin Alumni Research Foundation (T.H. and K.Z.) and by NSFC, NCET and HuoYingDong Foundation (Z.S.).  M.J.S is supported by the US DOE under DE-FG02-96DR40949.
}
  
\bibliography{hidvalley}
\bibliographystyle{JHEP}

\end{document}